\documentclass[aps,prd,showpacs,onecolumn,preprintnumbers,nofootinbib,10pt]{revtex4-1}
\usepackage{amsmath,bm}
\usepackage{amsfonts}
\usepackage{amssymb}
\usepackage{braket}
\usepackage{color,graphicx}
\usepackage{epstopdf}
\usepackage[dvipsnames]{xcolor}
\usepackage{slashed}
\usepackage{float}
\usepackage{multirow}
\usepackage{hyperref}
\hypersetup{colorlinks,citecolor=nicegreen,linkcolor=nicered,urlcolor=RoyalBlue}
\usepackage{enumitem}
\usepackage[left=2.0cm,right=2.0cm,top=2.0cm,bottom=2.5cm]{geometry}

\newcommand{\beq}{\begin{eqnarray}}
\newcommand{\eeq}{\end{eqnarray}}
\newcommand{\non}{\nonumber\\ }

\newcommand{\psl}{ P \hspace{-2.4truemm}/ }

\newcommand{\epsl}{\epsilon \hspace{-1.6truemm}/\,  }

\def\lsim{ {\ \lower-1.2pt\vbox{\hbox{\rlap{$<$}\lower6pt\vbox{\hbox{$\sim$}
}}}\ } }
\def\gsim{ {\ \lower-1.2pt\vbox{\hbox{\rlap{$>$}\lower6pt\vbox{\hbox{$\sim$}
}}}\ } }

\def \jhep{ J. High Energy Phys.  }

\definecolor{Red}{rgb}{1.,0.,0.}
\definecolor{Blue}{rgb}{0.,0.,1.}
\definecolor{RoyalBlue}{rgb}{0.0,0.14,0.4}
\definecolor{nicered}{rgb}{0.7,0.1,0.2}
\definecolor{nicegreen}{rgb}{0.1,0.4,0.2}

\def\orcid#1{\kern .08em\href{https://orcid.org/#1}{\includegraphics[keepaspectratio,width=0.76em]{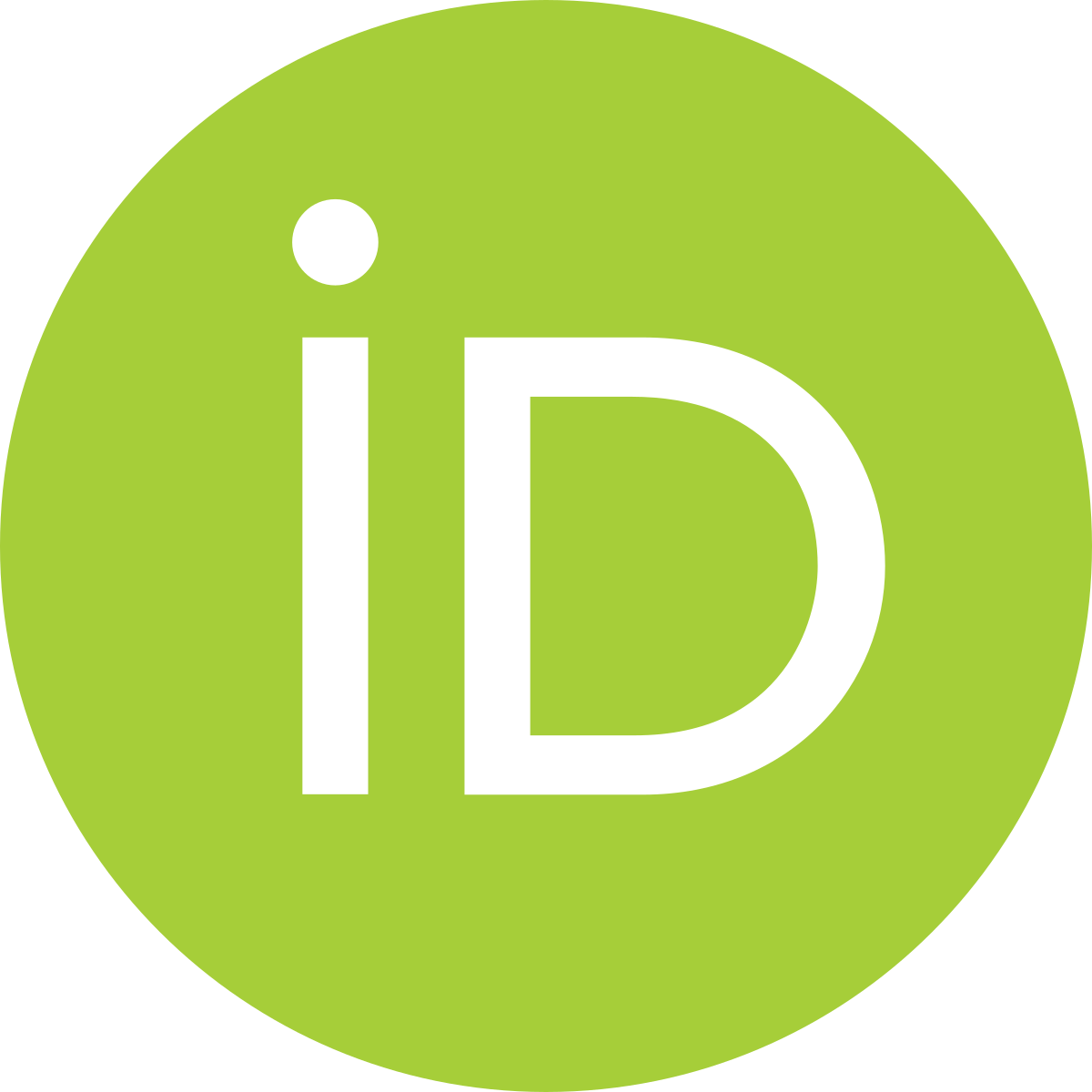}}}

\begin{document}

\title{\boldmath Phenomenological studies on neutral $B$-meson decays
into $J/\psi f_1$ and $\eta_c f_1$}
\author{De-Hua~Yao}
\author{Xin~Liu\orcid{0000-0001-9419-7462}}
\email [ Corresponding author: ]
{liuxin@jsnu.edu.cn}
\affiliation{Department of Physics,
Jiangsu Normal University, Xuzhou 221116, China}

\author{Zhi-Tian~Zou}
\author{Ying~Li\orcid{0000-0002-1337-7662}}
\email [ Corresponding author: ]
{liying@ytu.edu.cn}
\affiliation{ Department of Physics, Yantai University, Yantai 264005,
 China}

\author{Zhen-Jun~Xiao\orcid{0000-0002-4879-209X}}
\affiliation{Department of Physics and Institute of Theoretical Physics,\\
Nanjing Normal University, Nanjing 210023, China}

\date{\today{}}

\begin{abstract}

The axial-vector mesons $f_1(1285)$ and $f_1(1420)$ are particularly viewed as the mixtures of flavor states $f_n$ and $f_s$ with mixing angle $\varphi$. In order to determine this angle, we study the $B_{d,s}^0\to J/\psi f_1(1285,1420)$ and $B_{d,s}^0\to \eta_c f_1(1285,1420)$ decays in perturbative QCD (PQCD) approach, including the effects of vertex corrections, nonfactorizable diagrams and penguin operators. Not only the branching fractions, but also the direct $CP$ asymmetries and the polarization fractions are calculated. It is found that the branching fractions of these decays are large enough to be measured in the running LHCb and Belle-II experiments. Moreover, in comparison with the observed ${\cal B}(B_{d,s}^0 \to J/\psi f_1(1285))$, $B_s^0 \to (J/\psi, \eta_c) f_1(1420)$ decays have large branching fractions, which could be measured promisingly through $f_1(1420) \to K_S^0 K^\pm \pi^\mp$ in experiments. We also propose several ratios that could be used to further constrain the absolute value of the mixing angle $\varphi$, but its sign cannot be determined yet in these decays. The direct CP asymmetries of these decays indicate the penguin pollution in the $B_d^0 \to (J/\psi, \eta_c) f_1$ decays cannot be neglected. We acknowledge that there are large theoretical uncertainties arising from the distribution amplitudes of axial-vector mesons and charmonium states, and more precise nonperturbative parameters are called. The comparisons between our results and future experimental data would help us to understand the nature of $f_1$ states and to test the PQCD approach.

\end{abstract}

\pacs{13.25.Hw, 12.38.Bx, 14.40.Nd}
\preprint{\footnotesize  JSNU-PHY-HEP-01/22}
\maketitle


\newpage
%
%
\section{Introduction}
\label{sect:1}

In the quark model, two nonets of $J^P = 1^+$ axial-vector mesons are expected as the orbital excitation of the $\bar q q$ system. In terms of the spectroscopic notation $^{2S+1}L_J$, there are two types of $P$-wave axial-vector mesons, namely, $^{3}P_{1}$ and $^{1}P_{1}$. These two nonets have distinctive $C$ quantum numbers for the corresponding neutral mesons, $C=+$ and $C=-$, respectively \cite{Cheng:2011pb, Burakovsky:1997dd, Chen:2015iqa}. The light axial-vector $f_1$ states, namely, $f_1(1285)$ and $f_1(1420)$, accompanied with $a_{1}(1260)$ and $K_{1A}$, are categorized as the $1^{++}$ multiplets, while the $1^{+-}$ multiplets incorporate $b_{1}(1235), h_{1}(1170), h_{1}(1380)$ and $K_{1B}$~\cite{ParticleDataGroup:2022pth, HFLAV:2022pwe}.  Although lots of efforts have been made to investigate these light axial-vectors \cite{Du:2022nno, Gidal:1987bn,Close:1997nm,Li:2000dy, Carvalho:2002fh, Li:2005eq, Yang:2007zt, Cheng:2007mx,  Yang:2008xw,Cheng:2008gxa, Yang:2010ah, Dudek:2011tt, Stone:2013eaa, Dudek:2013yja, Liu:2014doa, Liu:2014jsa,  Close:2015rza, Molina:2016pbg, Liu:2016rqu,  Jiang:2020eml,He:2021exv}, our understanding on their natures is still far from satisfactory \cite{Du:2022nno}. Similar to the $\eta-\eta^\prime$ mixing in the pseudoscalar sector, two physical $f_1$ mesons (For convenience, we will adopt $f_1$ to denote the $f_1(1285)$ and $f_1(1420)$ mesons in the following context, unless otherwise stated.) are generally viewed as the mixtures of two flavor states $f_n( \equiv (\bar u u+ \bar d d)/{\sqrt{2}})$ and $f_s( \equiv\bar s s)$ with a single mixing angle $\varphi$, which can be written as
\beq
\left(
\begin{array}{c} f_1(1285)\\ f_1(1420) \\ \end{array} \right ) &=&
\left( \begin{array}{cc}
\cos{\varphi} & -\sin{\varphi} \\
\sin{\varphi} & \; \;\ \cos{\varphi} \end{array} \right )
\left( \begin{array}{c}  f_{n}\\ f_{s} \\ \end{array} \right )\;.
\label{eq:mix-fn-fs}
\eeq
However, both the magnitude and the sign of the mixing angle $\varphi$ have not been determined yet. In addition, in light of Gell-mann$-$Okubo mass relation, this mixing angle $\varphi$ could provide constraints to the unique mixing between the $K_{1A}(1^3P_1)$ and $K_{1B}(1^1P_1)$ states for the axial-vector strange $K_1$ mesons\cite{Cheng:2011pb}. In order to study the mixing angle $\varphi$, besides the $f_1$ mesons decays, the productions of $f_1$ mesons in the nonleptonic decays of heavy mesons could also be used. With this strategy, many $B$ decays to $f_1$ have been explored in the literature~\cite{Cheng:2007mx, Yang:2008xw, Liu:2014doa, Liu:2014jsa, Liu:2016rqu, Jiang:2020eml, Close:2015rza, Molina:2016pbg, He:2021exv}.

In 2013, LHCb collaboration reported their first measurements of the branching fractions of $B_{d,s}^0 \to J/\psi f_1(1285)$ decays as follows~\cite{Aaij:2013rja},
\beq
&&{\cal B}(B_d^0 \to J/\psi f_1(1285))_{\rm Exp}=(8.37^{+2.10}_{-2.09}) \times 10^{-6}, \label{eq:psif12d-ex}\\
&&{\cal B}(B_s^0 \to J/\psi f_1(1285))_{\rm Exp}=(7.14^{+1.36}_{-1.41}) \times 10^{-5}, \label{eq:psif12s-ex}
\eeq
where the uncertainties from different sources have been added in quadrature. It is found that the uncertainties are still large, and are expected to be reduced in the on-going LHCb and Belle-II experiments. Using the SU(3) symmetry and neglecting the contributions from penguin operators, the mixing angle $\varphi$ was extracted to be $\varphi^{\rm Exp}= \pm (24.0^{+3.1+0.6} _{-2.6-0.8})^\circ$ \cite{Aaij:2013rja}. In Ref.\cite{Liu:2014doa}, two of us (Liu and Xiao) had studied $B_s^0 \to J/\psi f_1(1285)$ decay and obtained $|\varphi^{\rm Theo}| \sim 15^\circ$, where the large errors arose from large theoretical uncertainties of the branching fraction. Although both $|\varphi^{\rm Theo}|$ and $|\varphi^{\rm Exp}|$ locate in the range proposed by Stone and Zhang~\cite{Stone:2013eaa}, the evident discrepancy still demands further explorations.

\begin{figure}[!!t]
\centering
\begin{tabular}{l}
\includegraphics[width=0.85\textwidth]{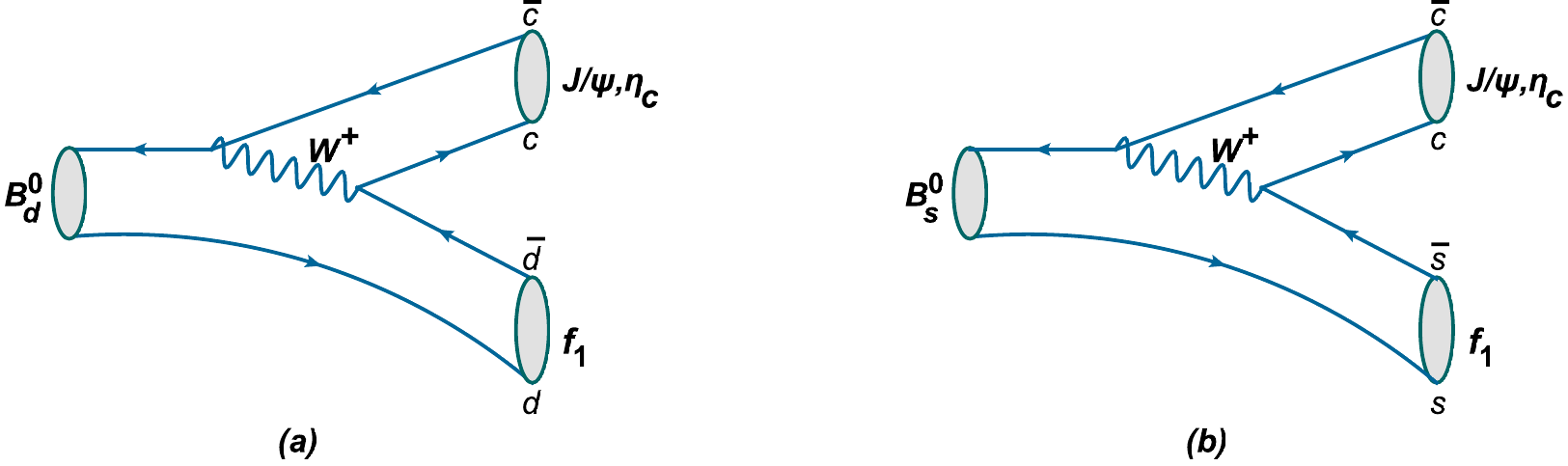}
\end{tabular}
\caption{(Color online)  Leading quark-level Feynman diagrams
for neutral $B$-meson decays into $J/\psi f_1$  and $\eta_c f_1$.
}
  \label{fig:fig1}
\end{figure}

It is the purpose of this article to analyze the $B_{d,s}^0 \to J/\psi f_1$ decays and to search for new observables for determination of the mixing angle, and the quark level Feynman diagrams for these decays are illustrated in Figure.\ref{fig:fig1}. The $B$-meson decays with a charmonium state have been studied extensively in the QCD-inspired approaches ~\cite{Cheng:2000kt, Cheng:2001ez, Song:2002gw, Meng:2005er, Chen:2005ht, Li:2006vq, Li:2007xf, Beneke:2008pi, Liu:2009yno, Colangelo:2010wg, Liu:2012ib, Li:2012sw,Wang:2015uea, Liu:2019ymi, Liu:2013nea, Xiao:2019mpm}. In this article we focus on the perturbative QCD (PQCD) approach to non-leptonic $B$-decays \cite{Keum:2000ph,Lu:2000em,Ali:2007ff}. It is a model-independent framework that systematically disentangles short-distance (perturbative) from long-distance (non-perturbative) effects based on the $k_T$ factorization, and the basic concepts will be given in next section. It is found that for the color-suppressed modes, the contributions beyond leading order (LO) play important roles in explaining the experimental data. For instance, $B \to J/\psi V$ and $B \to \eta_c V$ decays have been studied in PQCD approach in Refs.~\cite{Xiao:2019mpm, Liu:2013nea} associated with the next-to-leading order (NLO) contributions, namely, the vertex corrections and the NLO Wilson coefficients, and the theoretical predictions are improved to basically agree with current data~\cite{ParticleDataGroup:2022pth, HFLAV:2022pwe}. Recently, the Sudakov factor for charmonium that plays critical roles in suppressing the long-distance contributions was  derived, which affects the observables of the decays with charmonium remarkably \cite{Liu:2018kuo, Liu:2020upy}. With above new ingredients, we will reexamine the decays $B_s^0 \to J/\psi f_1$~\cite{Liu:2014doa} and evaluate the modes $B_d^0 \to J/\psi f_1$ and $B^0 \to \eta_c f_1$ for the first time, and will provide predictions of the branching fractions,  polarization fractions, relative phases and CP asymmetries. Within the experimental data, the branching fractions could help us constrain $|\varphi|$ effectively, while other observables are helpful to further understand the QCD  of $f_1$. Moreover, as pointed out in Refs.~\cite{Yang:2007zt,  Cheng:2007mx, Yang:2008xw, Cheng:2008gxa}, the QCD behavior of $1^{++}$ axial-vector meson is very similar to that of $1^{--}$ vector, then it is natural to expect some important information provided by the considered $B_s^0$ decay modes, relative to the golden channel $B_s^0 \to J/\psi \phi$. For example, $B_s^0 \to J/\psi f_1(1420)$ and $B_s^0 \to \eta_c f_1(1420)$ decays might serve as the alternative channels to explore the $B_s^0-\bar B_s^0$ mixing phase $\phi_s$ in a supplementary manner, provided that the mixing angle has been well determined by other ways. After all, it is an unarguable fact that the exclusive decays of neutral $B$-meson into charmonium have attracted great attention in the past decades at both theoretical and experimental aspects,  as they can play special roles in studies of CP asymmetries \cite{Abe:2001oa} and $B^0-\bar B^0$ mixing phases.

The  paper is organized as follows. In Sect.~\ref{sect:2}, we shortly review the formalism of PQCD approach in association with the meson wave functions,  and then present the perturbative calculations of considered decays. The analytic expressions of decay amplitudes are also collected in this section. In Sect.~\ref{sect:3}, we perform the numerical evaluations and discuss the theoretical results. Finally, a brief summary of this work is given in Sect.~\ref{sect:4}.

%
%
\section{Formalism and Perturbative QCD calculations}
\label{sect:2}

We consider the $B$ meson at rest for simplicity. For the decays $B^0 \to M_{c\bar c} f_1$ with $M_{c\bar c}$ denoting the $J/\psi$ and $\eta_c$, $M_{c\bar c}$ and $f_{1}$ are assumed to move in the plus and minus $z$-directions, respectively. In the light-cone coordinate, the $B$ meson momentum $P_1$, the charmonium momentum $P_2$ and $f_1$ meson momentum $P_3$ are taken to be:

\beq
P_{1}=\frac{m_{B}}{\sqrt{2}}(1, 1, {\bf 0}_{T})\;,
\qquad
P_{2}= \frac{m_{B}}{\sqrt{2}}(1-r_{3}^{2}, r_{2}^{2}, {\bf 0}_{T})\;,
\qquad
P_{3}=\frac{m_{B}}{\sqrt{2}}(r_{3}^{2}, 1-r_{2}^{2}, {\bf 0}_{T})\;,
\label{eq:coor2}
\eeq
and the polarization vectors $\epsilon_2$ of $J/\psi$ and $\epsilon_3$ of $f_1$ as,
\beq
\epsilon_{2L}&=& \frac{1}{\sqrt{2(1-r_{3}^{2})} r_{2}}(1-r_{3}^{2},
-r_{2}^{2}, {\bf 0}_{T}),
\qquad
\epsilon_{2T}=(0,0, {\bf 1}_{T})  \;,
\label{eq:polvec-psi}
\eeq
and
\beq
\epsilon_{3L}&=&\frac{1}{\sqrt{2(1-r_{2}^{2})} r_{3}}(-r_{3}^{2}, 1-r_{2}^{2}, {\bf 0}_{T}),
\qquad
\epsilon_{3T}=(0,0, {\bf 1}_{T})\;.
\label{eq:polvec-f1}
\eeq
where the ratios $r_{2}=m_{M_{c\bar c}}/m_{B}$ and $r_{3}=m_{f_{1}}/m_{B}$. Due to the conservation of the  angular momentum, only the longitudinal polarization vector $\epsilon_{3L}$ of $f_1$  contributes to the $B^0 \to \eta_c f_1$ decays. Denoting the (light-)quark momenta in the $B^0$, $M_{c\bar c}$ and $f_{1}$ mesons as $k_{1}$, $k_{2}$ and $k_{_{3}}$ correspondingly, we then have
\beq
k_{1} &=& (x_{1}P_{1}^{+}, 0, {\bf k}_{1T})
= (\frac{m_{B}}{\sqrt{2}} x_{1}, 0, {\bf k}_{1T})\;,
\label{eq:lqm-b} \nonumber
\\
k_{2} &=&
(x_{2}P_{2}^{+}, x_{2}P_{2}^{-}, {\bf k}_{2T})
= (\frac{m_{B}}{\sqrt{2}} x_{2}(1-r_{3}^{2}),
\frac{m_{B}}{\sqrt{2}} x_{2} r_{2}^{2}, {\bf k}_{2T})\;,
\label{eq:lqm-mcc}
\\
k_{3} &=&
(x_{3}P_{3}^{+}, x_{3}P_{3}^{-}, {\bf k}_{3T})
= (\frac{m_{B}}{\sqrt{2}} x_{3}r_{3}^{2},
\frac{m_{B}}{\sqrt{2}} x_{3}(1-r_{2}^{2}), {\bf k}_{3T})\;.\nonumber
\label{eq:lqm-fq}
\eeq

For the considered decays $B^0 \to M_{c\bar c} f_1$, the effective Hamiltonian $H_{\rm eff}$ could be read as~\cite{Buchalla:1995vs}
\beq
H_{\rm eff}\, &=&\, \frac{G_F}{\sqrt{2}}
\biggl\{ V^*_{cb}V_{cq} \biggl[ C_1(\mu)O_1^{c}(\mu)
+C_2(\mu)O_2^{c}(\mu) \biggr]
- V^*_{tb}V_{tq} \biggl[ \sum_{i=3}^{10}C_i(\mu)O_i(\mu) \biggr] \biggr\}\;,
\label{eq:heff}
\eeq
where the light quark $q=d$ or $s$, the Fermi constant $G_F=1.16639\times 10^{-5}{\rm GeV}^{-2}$, $V_{ij}$ represents the Cabibbo-Kobayashi-Maskawa (CKM) matrix element, and $C_i(\mu)$ is Wilson coefficients corresponding to the effective operator $O_i$ at the renormalization scale $\mu$. The local four-quark operators $O_i(i=1,\cdots,10)$ are given as
\begin{itemize}
\item Tree operators
\begin{eqnarray}
\begin{array}{ll}
\displaystyle
O_1^{c}\, =\,
(\bar{q}_\alpha c_\beta)_{V-A}(\bar{c}_\beta b_\alpha)_{V-A}\;,
& \displaystyle
O_2^{c}\, =\, (\bar{q}_\alpha c_\alpha)_{V-A}(\bar{c}_\beta b_\beta)_{V-A}\;,
\label{eq:operators-1}
\end{array}
\end{eqnarray}
\item  QCD penguin operators
\begin{eqnarray}
\begin{array}{ll}
\displaystyle
O_3\, =\, (\bar{q}_\alpha b_\alpha)_{V-A}\sum_{q'}(\bar{q}'_\beta q'_\beta)_{V-A}\;,
& \displaystyle
O_4\, =\, (\bar{q}_\alpha b_\beta)_{V-A}\sum_{q'}(\bar{q}'_\beta q'_\alpha)_{V-A}\;,
\\
\displaystyle
O_5\, =\, (\bar{q}_\alpha b_\alpha)_{V-A}\sum_{q'}(\bar{q}'_\beta q'_\beta)_{V+A}\;,
& \displaystyle
O_6\, =\, (\bar{q}_\alpha b_\beta)_{V-A}\sum_{q'}(\bar{q}'_\beta q'_\alpha)_{V+A}\;,
\end{array}
\label{eq:operators-2}
\end{eqnarray}
\item  Electroweak penguin operators
\begin{eqnarray}
\begin{array}{ll}
\displaystyle
O_7\, =\,
\frac{3}{2}(\bar{q}_\alpha b_\alpha)_{V-A}\sum_{q'}e_{q'}(\bar{q}'_\beta q'_\beta)_{V+A}\;,
& \displaystyle
O_8\, =\,
\frac{3}{2}(\bar{q}_\alpha b_\beta)_{V-A}\sum_{q'}e_{q'}(\bar{q}'_\beta q'_\alpha)_{V+A}\;,
\\
\displaystyle
O_9\, =\,
\frac{3}{2}(\bar{q}_\alpha b_\alpha)_{V-A}\sum_{q'}e_{q'}(\bar{q}'_\beta q'_\beta)_{V-A}\;,
& \displaystyle
O_{10}\, =\,
\frac{3}{2}(\bar{q}_\alpha b_\beta)_{V-A}\sum_{q'}e_{q'}(\bar{q}'_\beta q'_\alpha)_{V-A}\;,
\end{array}
\label{eq:operators-3}
\end{eqnarray}
\end{itemize}
where $\alpha$ and $\beta$ are the color indices and $q^\prime$ are the active quarks at the scale $m_b$, i.e.  $q^\prime=(u,d,s,c,b)$. The left handed current is defined as $({\bar{q}}^{\prime}_{\alpha} q^{\prime}_{\beta} )_{V-A}= {\bar{q}}^{\prime}_{\alpha} \gamma_\nu (1-\gamma_5) q^{\prime}_{\beta}$ and the right handed current $({\bar{q}}^{\prime}_{\alpha} q^{\prime}_{\beta} )_{V+A}= {\bar{q}}^{\prime}_{\alpha} \gamma_\nu (1+\gamma_5) q^{\prime}_{\beta}  $. For convenience, the combination $a_{i}$ of the Wilson coefficients is defined as~\cite{Ali:2007ff,Zou:2015iwa}
\begin{eqnarray}
a_1= C_2+C_1/3, & a_3= C_3+C_4/3,~a_5= C_5+C_6/3,~a_7=
C_7+C_8/3,~a_9= C_9+C_{10}/3,\nonumber \\
 a_2= C_1+C_2/3, & a_4=
C_4+C_3/3,~a_6= C_6+C_5/3,~a_8= C_8+C_7/3,~a_{10}= C_{10}+C_{9}/3.
\end{eqnarray}

Under the factorization hypothesis, the amplitude of $B^0 \to M_{c\bar c} f_1$ decay in PQCD can be written conceptually as follows  \cite{Keum:2000ph,Lu:2000em,Ali:2007ff},
\beq
{\cal A}(B^0 \to M_{c\bar c} f_1) &\sim &\int\!\! d x_1 d
x_2 d x_3 b_1 d b_1 b_2 d b_2 b_3 d b_3
\non &\times & {\rm Tr}
\left [ C(t) \Phi_{B}(x_1, b_1) \Phi_{M_{c\bar c}}(x_2, b_2)
\Phi_{f_1}(x_3, b_3) H(x_i, b_i, t) S_t(x_i)\, e^{-S(t)} \right ]\;,
\label{eq:amp-f1}
\eeq
where $C(t)$ stands for the corresponding Wilson coefficient. $x_{i}(i=1,2,3)$ denotes the fraction of momentum carried by (light-)quark inside the meson, and $b_{i}$ is the conjugate space coordinate of transverse momentum $k_{iT}$. The wave function $\Phi$ describes the hadronization of quark and antiquark into a meson, and it is nonperturbative in nature but universal. The hard kernel $H(x_i, b_i, t)$ involves the  dynamics associated with the effective ``six-fermion interaction'' by exchanging a hard gluon~\cite{Keum:2000wi,Lu:2000em, Keum:2000ph, Lu:2002ny}, where $t$ is the largest energy scale involved in the hard part. The rest two factors, namely, the Sudakov factor $e^{-S(t)}$ and the jet function $S_t(x_i)$ as shown in the above Eq.~(\ref{eq:amp-f1}), play important roles on the effective evaluations of $B$-meson decay amplitude in PQCD, which is based on the $k_T$ factorization theorem. The Sudakov factor $e^{-S(t)}$ suppresses the soft dynamics effectively, which makes perturbative calculation of the hard part $H$ applicable at intermediate scale~\cite{Botts:1989nd, Li:1992nu}. The jet function $S_t(x_i)$ smears the endpoint singularities with threshold resummation technique~\cite{Li:2001ay,Li:2002mi}. Recently, the Sudakov factor for charmonium up to the next-to-leading-logarithm accuracy has been derived in Refs.~\cite{Liu:2018kuo, Liu:2020upy}, where the effects of the charm quark mass are also included. In this work, we will adopt the new factor. Besides, more concepts of PQCD can be found in Ref.~\cite{Li:2003yj}. In recent years, several developments on this approach have been obtained, for a review, see, e.g.~\cite{Cheng:2020fcx,Hua:2020usv}.

\subsection{Meson Wave Functions}

As aforementioned, the nonperturbative meson wave functions and the related distribution amplitudes are the most important inputs in PQCD, which are universal and usually determined within the experimental data or the nonperturbative techniques, such as QCD sum rule or Lattice QCD.

For the $B$ meson wave function, we adopt the form used widely in the literature~\cite{Keum:2000ph, Keum:2000wi, Lu:2000em, Lu:2002ny, Li:2003yj}, which is expressed as
\beq
\Phi_{B}(x, {\bf k}_T) &=& \frac{i }{\sqrt{2N_c}} \biggl\{(\psl_{1} +m_{B})\gamma_5
\phi_{B}(x, {\bf k}_T) \biggr\}_{\alpha\beta},
\label{eq:wf-B}
\eeq
$N_c=3$ being the color factor. $\phi_B$ is the leading-twist $B$-meson distribution amplitude, and $x$ and ${\bf k}_T$ are the momentum fraction and the intrinsic transverse momentum of light quark in $B$ meson, respectively. The subscripts $\alpha$ and $\beta$ are the color indices.

For charmonium states $J/\psi$ and $\eta_c$, their wave functions have been studied within the non-relativistic QCD approach \cite{Bondar:2004sv}. For the vector $J/\psi$ meson, the longitudinal and transverse wave functions are given as,
\beq
\Phi_{J/\psi}^{L}(x) &=& \frac{1}{\sqrt{2N_{c}}}
\biggl\{m_{J/\psi}\epsl_{L}\phi_{J/\psi}^{L}(x)
+\epsl_{L}\psl_2\ \phi_{J/\psi}^{t}(x)\biggl\}_{\alpha\beta}\;,
\label{eq:wf-psi-L}\\
\Phi_{J/\psi}^{T}(x) &=& \frac{1}{\sqrt{2N_{c}}}
\biggl\{m_{J/\psi}\epsl_{T}\phi_{J/\psi}^{v}(x)
+\epsl_{T}\psl_2\ \phi_{J/\psi}^{T}(x)\biggl\}_{\alpha\beta}\;.
\label{eq:wf-psi-T}
\eeq
Here, $\epsilon_{L}$ and $\epsilon_{T}$ are the two polarization vectors of $J/\psi$, $\phi^{L}_{J/\psi}(x)$ and $\phi^{T}_{J/\psi}(x)$ are the twist-2 distribution amplitudes, while $\phi^{t}_{J/\psi}(x)$ and $\phi^{v}_{J/\psi}(x)$ are the twist-3 ones.  For the pseudoscalar $\eta_{c}$ meson, its wave function could be read as,
\beq
\Phi_{\eta_{c}}(x) &=& \frac{i}{\sqrt{2N_{c}}}\gamma_{5}
\biggl\{\psl \phi_{\eta_{c}}^{v}(x)+m_{\eta_{c}}\phi_{\eta_{c}}^{s}(x)
\biggl\}_{\alpha\beta}\;,
\label{eq:wf-etac}
\eeq
where $\phi_{\eta_{c}}^{v}(x)$ and $\phi_{\eta_{c}}^{s}(x)$ are the twist-2 and twist-3 distribution amplitudes, respectively.

For the light axial-vector $f_{q}$ with $q=n$ or $s$, its wave function could be written as follows~\cite{Yang:2007zt,Li:2009tx},
\beq
\Phi^{L}_{f_q}(x) &=&  \frac{1}{\sqrt{2 N_c}}\gamma_5
\biggl\{ m_{f_q}\, {\epsl}_L \,\phi_{f_q}(x)  +
{\epsl}_L \, \psl_3\,\phi_{f_q}^t(x)  + m_{f_{q}}\, \phi_{f_q}^s(x) \biggr\}_{\alpha\beta}\;,
\label{eq:wf-f1-L}
\\
\Phi^{T}_{f_q}(x) &=&  \frac{ 1}{\sqrt{2 N_c}} \gamma_5
\biggl\{ m_{f_q}\, {\epsl}_T\, \phi_{f_q}^v(x) +
{\epsl}_T\, \psl_3\, \phi_{f_q}^T(x)+m_{f_q}
i\epsilon_{\mu\nu\rho\sigma}\gamma_5\gamma^\mu \epsilon_T^{\nu}
n^\rho v^\sigma \phi_{f_q}^a(x) \biggr\}_{\alpha\beta}\;,
\label{eq:wf-f1-T}
\eeq
where $x$ denotes the momentum fraction carried by quark in $f_q$, $n=(1,0,{\bf 0}_T)$ and $v=(0,1,{\bf 0}_T)$ are the dimensionless lightlike vectors, the Levi-Civit$\grave{a}$ tensor $\epsilon^{\mu\nu\alpha\beta}$ is conventionally taken as $\epsilon^{0123}=1$. It  should be stressed that $m_{f_q}$ stands for the mass of $f_q$ obtained through the following mass relations~\cite{Verma:2011yw},
\beq
m_{f_n}^2 &=& m_{f_1(1285)}^2 \cos^2\varphi + m_{f_1(1420)}^2 \sin^2\varphi\;,\\
m_{f_s}^2 &=& m_{f_1(1285)}^2 \sin^2\varphi + m_{f_1(1420)}^2 \cos^2\varphi\;,
\label{eq:mass-fs}
\eeq
with $m_{f_1(1285)}$ and $m_{f_1(1420)}$ being the masses of physical states $f_1(1285)$ and $f_1(1420)$, respectively.

For the sake of simplicity, the distribution amplitudes in the above wave functions have been collected in Appendix.~\ref{sec:app0}. With the above essential hadronic inputs, we can then proceed with perturbative calculations of the $B^0 \to M_{c\bar c} f_1$ decay amplitudes within the framework of PQCD approach.

\subsection{ Perturbative calculations
}

\begin{figure}[!!htb]
\begin{center}
\includegraphics[scale=0.95]{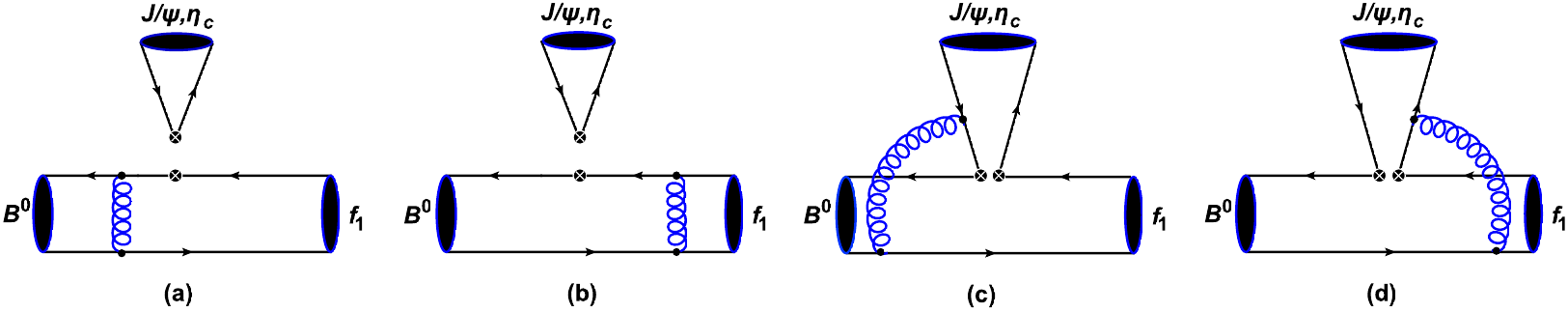}
\caption{(Color online)  Typical Feynman diagrams
 for neutral $B$-meson decays into $J/\psi f_{1}$ and $\eta_c f_1$
at LO in the PQCD approach.  }
\label{fig:fig2}
\end{center}
\end{figure}

It can be seen from  Figure.~\ref{fig:fig1} that for the $B_d^0 \to M_{c\bar c} f_1$ decays, the spectator $d$ quark enters the final axial vector meson $f_1$, therefore only $f_n$ component contributes to the decays with mixing angles. Similarly, only $f_s$ component contributes to the $B_s^0 \to M_{c\bar c} f_1$ decays. According to the effective Hamiltonian eq.~(\ref{eq:heff}), the lowest-order (LO) Feynman diagrams are summarized in Figure.~\ref{fig:fig2} for $B \to M_{c\bar c} f_1$ decays, where the first two diagrams are called factorizable and the last two diagrams are the non-factorizable diagrams. Due to the fact that the behavior of axial vector in $B \to M_{c\bar c} f_1$ decays is very similar to that of vectors, the amplitudes $F$ for the factorizable diagrams and $M$ for the nonfactorizable ones are almost same as those of decays $B \to (J/\psi, \eta_c) V$ decays. In this case, we will not list them in current work, and the readers are referred to Refs.~\cite{Liu:2013nea,Xiao:2019mpm} for detail. It should be stressed that the term $\sqrt{1-r_2^2}$ in the denominator of the longitudinal polarization vector $\epsilon_{3L}$ is kept in Eq.~(\ref{eq:polvec-f1}), while it has been neglected in Ref.~\cite{Liu:2013nea}.

\begin{figure}[!!htb]
\begin{center}
\includegraphics[scale=0.95]{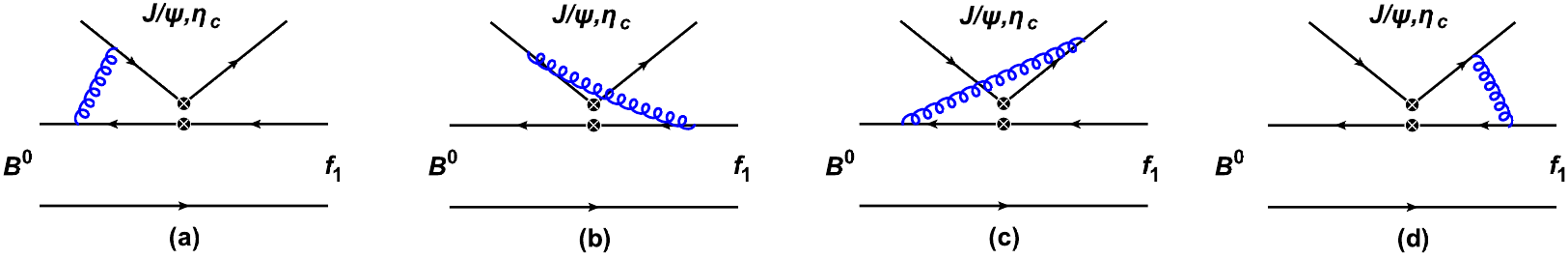}
\caption{(Color online)  Vertex corrections to
neutral $B$-meson decays into $J/\psi f_{1}$ and $\eta_c f_1$.}
\label{fig:fig3}
\end{center}
\end{figure}

In Refs.\cite{Cheng:2000kt,Cheng:2001ez,Song:2002gw,Meng:2005er,Chen:2005ht,Li:2006vq,Li:2007xf,Beneke:2008pi,Colangelo:2010wg}, it was found that for the color-suppressed processes, such as $B$-meson decays into charmonium states, the vertex corrections play so important roles in explaining the experimental data that cannot be neglected. For this reason, the vertex corrections in $B^0 \to M_{c\bar c} f_1$ decays as illustrated in Figure.~\ref{fig:fig3} should be included, and their effects are embodied by modifying the Wilson coefficients in the factorizable emission diagrams, leading to the effective Wilson coefficients $\tilde{a}_i^h (i=2,3,5,7,9)$ as follows,
\beq
a_2 &\to& \tilde{a}_{2}^{h} = a_2
+\frac{\alpha_{s}}{4\pi}\frac{C_{F}}{N_{c}}C_{2}
\biggl(-18+12\ln\frac{m_{b}}{\mu}+f_{I}^{h}\biggr)\;,
\label{eq:a2eff}
\\
a_3 &\to& \tilde{a}_{3}^{h} = a_3
+\frac{\alpha_{s}}{4\pi}\frac{C_{F}}{N_{c}}C_{4}
\biggl(-18+12\ln\frac{m_{b}}{\mu}+f_{I}^{h}\biggr)\;,
\label{eq:a3eff}
\\
a_5 &\to& \tilde{a}_{5}^{h} = a_5
+\frac{\alpha_{s}}{4\pi}\frac{C_{F}}{N_{c}}C_{6}
\biggl(6-12\ln\frac{m_{b}}{\mu}-f_{I}^{h}\biggr)\;,
\label{eq:a5eff}
\\
a_7 &\to& \tilde{a}_{7}^{h} = a_7
+\frac{\alpha_{s}}{4\pi}\frac{C_{F}}{N_{c}}C_{8}
\biggl(6-12\ln\frac{m_{b}}{\mu}-f_{I}^{h}\biggr)\;,
\label{eq:a7eff}
\\
a_9 &\to& \tilde{a}_{9}^{h} = a_9
+\frac{\alpha_{s}}{4\pi}\frac{C_{F}}{N_{c}}C_{10}
\biggl(-18+12\ln\frac{m_{b}}{\mu}+f_{I}^{h}\biggr)\;.
\label{eq:a9eff}
\eeq
In above formulae, the function $f_{I}^{h}$ with helicities
$h=0,\pm$ are
defined as:
\beq
f_{I}^{0}&=&f_{I}+g_{I}(1-r_{2}^{2})\;,
\qquad
f_I^{\pm}=f_{I}\;,
\eeq
where the explicit expressions for the functions $f_I$ and $g_I$ can be found in
 Ref.~\cite{Cheng:2001ez}.

We also note that in the LO calculations we used the LO Wilson coefficients $C_{i}(m_{W})$ and the LO renormalization group evolution matrix $U(t,m)^{(0)}$ for the Wilson coefficient associating with the LO running coupling $\alpha_{s}$,
\beq
\alpha_{s}(t)=\frac{4\pi}{\beta_{0}\ln [t^{2}/\Lambda_{\rm QCD}^{2}]},
\eeq
where $\beta_{0}=(33-2N_{f})/3$. In the NLO contributions, it is natural for us to adopt the NLO Wilson coefficients $C_{i}(m_{W})$ and the NLO renormalization group evolution matrix $U(t,m,\alpha)$ with the running coupling $\alpha_{s}(t)$ at two-loop~\cite{Buchalla:1995vs},
\beq
\alpha_{s}(t)=\frac{4\pi}{\beta_{0}\ln(t^{2}/\Lambda_{\rm QCD}^{2})}\cdot \biggl
\{1-\frac{\beta_{1}}
{\beta_{0}^{2}}\cdot \frac{\ln
[\ln(t^{2}/\Lambda_{\rm QCD}^{2})]}{\ln(t^{2}/\Lambda_{\rm QCD}^{2})}\biggr\},
\eeq
where $\beta_{1}=(306-38N_{f})/3$. For the hadronic scale $\Lambda_{\rm QCD}$, $\Lambda_{\rm QCD}^{(4)}=0.287$ GeV (0.326 GeV) could be arrived within $\Lambda_{\rm QCD}^{(5)}=0.225$ GeV for the LO (NLO) case. In addition, we set $\mu_{0}=1.0$ GeV \cite{Xiao:2008sw} as the lower cut-off for the hard scale $t$.

With the amplitudes of each diagrams in Figure.\ref{fig:fig2}, the amplitude of $B^0 \to J/\psi f_{q}$ decay could be written as
\beq
\xi A^{\sigma}(B_{d(s)}^0 \to J/\psi f_{n(s)}) &=&
F_{J/\psi}^{\sigma} \biggl\{ V_{cb}^* V_{cd(s)} \tilde{a}^\sigma_{2}
-V_{tb}^* V_{td(s)} \biggl( \tilde{a}^\sigma_{3}+\tilde{a}^\sigma_{5}
+\tilde{a}^\sigma_{7}+\tilde{a}^\sigma_{9}\biggr) \biggr\}
\non  &&
+M_{J/\psi}^\sigma \biggl\{V_{cb}^* V_{cd(s)} C_{2}
-V_{tb}^* V_{td(s)}
\biggl( C_{4}-C_{6}-C_{8}+C_{10}\biggr) \biggr\}\;,
\label{eq:DecAmp-jpsi}
\eeq
with $\xi=\sqrt{2}$ and  $\xi=1$ for $f_n$ and $f_s$, respectively. The superscript $\sigma (= L, N, T)$ denotes the helicity state of the final states. Combining above amplitude and the quark-flavor mixing scheme as shown in Eq.~(\ref{eq:mix-fn-fs}), we finally obtain the amplitudes of the $B^0 \to J/\psi f_{1}$ decays,
\begin{itemize}
\item[(1)]{For $B_d^0 \to J/\psi f_1$ decays,}
 \beq
{\cal A}^{\sigma}(B_{d}^{0}\to J/\psi f_{1}(1285))
&=&  A^{\sigma}(B_{d}^{0}\to J/\psi f_{n}) \cos \varphi \;,
\label{eq:DecAmp-psif12-d}
\\
{\cal A}^{\sigma}(B_{d}^{0}\to J/\psi f_{1}(1420))
&=&  A^{\sigma}(B_{d}^{0}\to J/\psi f_{n}) \sin \varphi \;,
\label{eq:DecAmp-psif14-d}
\eeq

\item[(2)]{For $B_s^0 \to J/\psi f_1$ decays,}
 \beq
{\cal A}^{\sigma}(B_{s}^{0}\to J/\psi f_{1}(1285))
&=& - A^{\sigma}(B_{s}^{0}\to J/\psi f_{s}) \sin \varphi \;,
\label{eq:DecAmp-psif12-s}
\\
{\cal A}^{\sigma}(B_{s}^{0}\to J/\psi f_{1}(1420))
&=& A^{\sigma}(B_{s}^{0}\to J/\psi f_{s}) \cos \varphi \;.
\label{eq:DecAmp-psif14-s}
\eeq
\end{itemize}

Now, we turn to the amplitudes of $B^0 \to \eta_c f_1$ decays. Similarly, the amplitudes of $B^0 \to \eta_{c} f_{1}$ decays at NLO could be obtained straightforwardly by replacing the information of vector $J/\psi$ with that of pseudoscalar $\eta_c$ in Eqs.~(\ref{eq:DecAmp-jpsi})-(\ref{eq:DecAmp-psif14-s}). Of course, only the longitudinal contributions of $f_1$ mesons contribute to  the $B^0 \to \eta_c f_1$ decays, because of the conservation of the angular momentum. Therefore, we have
\beq
\xi A(B_{d(s)}^{0} \to \eta_{c} f_{n(s)})
&=&
F_{\eta_c}
\biggl\{ V_{cb}^{\ast} V_{cd(s)}
\tilde{a}_{2}-V_{tb}^{\ast} V_{td(s)}
\biggl( \tilde{a}_{3}- \tilde{a}_{5}- \tilde{a}_{7}+ \tilde{a}_{9}\biggr) \biggr\}
\non &&
+M_{\eta_c} \biggl\{V_{cb}^{\ast} V_{cd(s)} C_{2} -V_{tb}^{\ast} V_{td(s)}
\biggl( C_{4}-C_{6}-C_{8}+C_{10}\biggr) \biggr\}\;.
\label{eq:DecAmp-etac}
\eeq
Here, the effective Wilson coefficients $\tilde{a}_i$ have included the related vertex corrections with new functions $f^\prime_I$ and $g^\prime_I$ that arise from $\eta_c$ emission~\cite{Song:2002gw}. The explicit expressions of $f^{(\prime)}_{I}$ and $g^\prime_I$ can be found in Refs.~\cite{Cheng:2001ez, Song:2002gw, Liu:2013nea, Xiao:2019mpm}. Also, the amplitudes of $B^0 \to \eta_c f_{1}$ decays could be read as follows,
\begin{itemize}
\item[(1)]{For $B_d^0 \to \eta_c f_1$ decays,}
\beq
{\cal A}(B_{d}^{0}\to \eta_c f_{1}(1285))
&=&   A(B_{d}^{0}\to \eta_c f_{n}) \cos \varphi \;,
\label{eq:DecAmp-etacf12-d}
\\
{\cal A}(B_{d}^{0}\to \eta_c f_{1}(1420))
&=&   A(B_{d}^{0}\to \eta_c f_{n}) \sin \varphi \;,
\label{eq:DecAmp-etacf14-d}
\eeq
\item[(2)]{For $B_s^0 \to \eta_c f_1$ decays,}
\beq
{\cal A}(B_{s}^{0}\to \eta_c f_{1}(1285))
&=& - A(B_{s}^{0}\to \eta_c f_{s}) \sin \varphi \;,
\label{eq:DecAmp-etacf12-s}
\\
{\cal A}(B_{s}^{0}\to \eta_c f_{1}(1420))
&=& A(B_{s}^{0}\to \eta_c f_{s}) \cos \varphi \;.
\label{eq:DecAmp-etacf14-s}
\eeq
\end{itemize}

As a matter of fact, using these decay channels, we can only extract the absolute value $|\varphi|$  of the mixing angle. In order to determine its sign, we have to resort to other decays that the interference between $f_n$ and $f_s$ is involved, such as $B \to M f_1$ with $M$ being the open-charmed and charmless mesons~\cite{Liu:2019ymi}.

%
%
\section{Numerical results and discussions}
\label{sect:3}

In this section, we will perform numerical calculations based on the given analytic expressions to estimate the experimental observables in the $B^0 \to M_{c\bar c} f_1$ decays, such as the branching fractions and the direct CP asymmetries. Furthermore, the results of the polarization fractions in the decays $B^0 \to J/\psi f_1$ are also discussed.

\subsection{Input parameters}
\label{ssec:inputs}

In the numerical calculations, the input parameters such as meson masses (GeV), decay constants (GeV) and $B$-meson lifetimes (ps) will be listed~\cite{ParticleDataGroup:2022pth,Yang:2007zt,Verma:2011yw}:
\beq
&&m_W = 80.41\;,
\quad
m_{B_d^0}= 5.28\;,
\quad
m_{B_s^0}= 5.37\;,
\quad
m_b = 4.8 \;,
\quad
m_{c} = 1.50  \;,
\nonumber
\\
&&f_{J/\psi} = 0.405\pm 0.014 \;,
\quad
f_{\eta_{c}}=0.42\pm0.05\;,
\quad
f_{f_{n}} = 0.193 ^{+0.043}_{-0.038} \;,
\quad
f_{f_{s}} = 0.230 \pm0.009\;,\nonumber
\\
&&m_{J/\psi}=3.097\;,
\quad
m_{f_{1}(1285)} = 1.28\;,
\quad
m_{f_{1}(1420)}= 1.43\;,
\quad
m_{\eta_{c}} = 2.98\;,\nonumber
\\
&&\tau_{B_d^0}= 1.519\;,
\quad
\tau_{B_s^0}= 1.515\;.
\label{eq:inputs}
\eeq
For CKM matrix elements, we adopt the Wolfenstein parameterization~\cite{Wolfenstein:1983yz}
and the updated parameters~\cite{ParticleDataGroup:2022pth}: $A=0.790$, $\lambda=0.22650$,
$\bar\rho=0.141_{-0.017}^{+0.016}$, $\bar\eta=0.357_{-0.011}^{+0.011}$, in which
$\bar \rho \equiv \rho ( 1- \frac{\lambda^2}{2})$ and $\bar \eta \equiv \eta (1- \frac{\lambda^2}{2})$.

\subsection{ 
\boldmath{$B \to J/\psi f_1$}
}
\label{ssec:jpsif1}

We firstly focus on the $B^0 \to J/\psi f_1$ decays. The branching fraction of $B^0 \to J/\psi f_{1}$ decay can be written as
\beq
{\cal B}(B^0 \to J/\psi f_1)
 = \tau_{B} \frac{G_{F}^{2}|\bf{p_c}|}{16 \pi m^{2}_{B} }
\sum_{\sigma=L,N,T} {\cal A}^{(\sigma)\dagger } {\cal A}^{(\sigma)}\;,
\label{eq:br-psif1}
\eeq
where $\tau_{B}$ is the lifetime of neutral $B$-meson, $|\bf{p_c}|\equiv |\bf{p_{2z}}|=|\bf{p_{3z}}|$ is the three-momentum of the two outgoing final states in the center-of-mass frame of $B$ meson and ${\cal A}^\sigma$ denotes the helicity amplitudes of $B^0 \to J/\psi f_1$ modes as given in Eqs.~(\ref{eq:DecAmp-psif12-d})-(\ref{eq:DecAmp-psif14-s}).

For the mixing angle $\varphi$, we take both $|\varphi^{\rm Theo}| \sim 15^\circ$ and $|\varphi^{\rm Exp}| \sim 24^\circ$ as typical values to predict the ${\cal B}(B^0 \to J/\psi f_1)$ and further discuss them phenomenologically. By employing the decay amplitudes and various inputs, the $CP$-averaged branching fractions in the PQCD approach at NLO with uncertainties are presented in Table.~\ref{tab:br-jpsif1}. We acknowledge that there are many uncertainties in our calculations, and four kinds of uncertainties are included here. The first uncertainties are from the wave function of $B$ meson, and we adopt the variations of shape parameters $\omega_{B_d^0} = 0.40 \pm 0.04$~GeV and $\omega_{B_s^0} = 0.50 \pm 0.05$~GeV for $B_d^0$ and $B_s^0$, respectively. The second uncertainties come from the nonperturbative parameters in the wave functions of the final states, such as decay constants and the Gegenbauer moments, which are given in Appendix.~\ref{sec:app0}. The third uncertainties arise from the higher-order and higher-power corrections, which are characterized by varying the running hard scale $t_{\rm max}$ with 20\%, namely, from $0.8t$ to $1.2t$ in the hard kernel. The last errors are induced by the variations of $|\varphi^{\rm Theo}|=(15.0 \pm 1.5)^\circ ~( |\varphi^{\rm Exp}| = (24.0^{+3.2}_{-2.7})^\circ )$, respectively. From this table, one finds that all theoretical results agree roughly with the available measurements in $2\sigma$ standard deviations. Additionally, the changes induced by the mixing angle $|\varphi|$ are plagued by uncertainties arising from other parameters. In this regard, in order to determine the mixing angle $|\varphi|$, more stringent constraints from other observables are required. We also note that the large branching fractions of the $B^0 \to J/\psi f_1$ decays are expected to be tested at LHCb and Belle-II experiments~\cite{Belle-II:2018jsg} in future.
\begin{table}[hbt]
\caption{ The CP-averaged branching ratios for neutral $B$-meson decays into $J/\psi f_1$ in the PQCD approach}
\label{tab:br-jpsif1}
 \begin{center}\vspace{-0.3cm}{
\begin{tabular}[t]{c|c|c}
\hline  \hline
  Decay Modes   &  $|\varphi^{\rm Theo}| \sim 15^\circ$
&  $|\varphi^{\rm Exp}| \sim 24^\circ$
\\
\hline \hline
  ${\cal B}(B_{d}^{0} \to J/\psi f_{1}(1285))$
&
$(3.05^{+0.73+1.88+0.09+0.04}_{-0.56-1.40-0.12-0.04}) \times 10^{-5}$
 &
$(2.73^{+0.65+1.68+0.08+0.11}_{-0.50-1.26-0.11-0.14})\times 10^{-5}$
 \\
 \hline
  ${\cal B}(B_{d}^{0} \to J/\psi f_{1}(1420))$
&
$(1.99^{+0.47+1.22+0.06+0.41}_{-0.37-0.91-0.08-0.37}) \times 10^{-6}$
 &
$(4.92^{+1.16+3.00+0.14+1.29}_{-0.91-2.26-0.20-1.00})\times 10^{-6}$
 \\
 \hline
  ${\cal B}(B_{s}^{0} \to J/\psi f_{1}(1285))$
&
$(0.89^{+0.25+0.36+0.04+0.19}_{-0.18-0.28-0.04-0.16})\times 10^{-4}$
&
$(2.21^{+0.61+0.88+0.09+0.58}_{-0.46-0.71-0.10-0.45})\times 10^{-4}$
 \\
 \hline
  ${\cal B}(B_{s}^{0} \to J/\psi f_{1}(1420))$
&
$(1.14^{+0.32+0.45+0.05+0.02}_{-0.23-0.35-0.05-0.01})\times 10^{-3}$
&
$(1.02^{+0.28+0.40+0.04+0.04}_{-0.21-0.32-0.04-0.05})\times 10^{-3}$
\\
 \hline \hline
\end{tabular}}
\end{center}
\end{table}

The decays $B_s^0 \to J/\psi f_1(1285,1420)$ have been investigated previously by two of us (Liu and Xiao) in Ref.\cite{Liu:2014doa} by including the vertex corrections. Comparing the new results with previous ones, we find all results agree with each other with uncertainties, and the acceptable differences are from the effects from the new ingredients in Sudakov factor. Furthermore, for comparison, we also take the decay $B_s^0 \to J/\psi f_1(1285)$ as an example and calculate its branching fractions for different values of $|\varphi|$ without vertex corrections and new ingredients in Sudakov factor, and the results are given as
\beq
{\cal B}(B_{s}^0 \to J/\psi f_1(1285))|_{\varphi\sim 15^\circ} &=&
(0.32^{+0.09+0.05+0.05+0.06}_{-0.07-0.05-0.05-0.06})\times 10^{-4}\;,\\
{\cal B}(B_{s}^0 \to J/\psi f_1(1285))|_{\varphi\sim 24^\circ} &=&
(0.78^{+0.22+0.12+0.12+0.21}_{-0.17-0.13-0.10-0.16})\times 10^{-4}\;.
\eeq
It is obvious that the vertex corrections can enhance the branching fractions remarkably, because the Wilson coefficient $\tilde a_2$ is much larger than $a_2$, which has been also shown in Refs.\cite{Cheng:2000kt,Chen:2005ht}. We also note that the uncertainties from the scale $t$ are  declined from $30\%$ to $5\%$, as we expected.

We also acknowledge that our NLO calculation are incomplete. In principle, the NLO contributions should contain both the vertex corrections and hard spectator scattering (HS) amplitudes. For the charmless $B$ meson decays, the NLO effects of HS have been explored partly \cite{Li:2012nk, Cheng:2014fwa,Cheng:2020fcx}. However, for the decays with charm quark, the calculations of NLO corrections of HS are very complicated because the new scale $m_c$ is involved, and we shall left it as our future work.

On the experimental side, $\varphi^{\rm Exp} = \pm (24.0^{+3.1+0.6} _{-2.6-0.8}) ^\circ$ was extracted through partial-wave analysis only with ${\cal B}(f_1(1285) \to 2\pi^+ 2\pi^-) = 0.109 \pm 0.006$~\cite{Aaij:2013rja}, by assuming the SU(3) flavor symmetry and neglecting the penguin contributions in $B_d^0 \to J/\psi f_1(1285)$ and $B_s^0 \to J/\psi f_1(1285)$ decays. As aforementioned, $f_1(1285)$ state could mix with $f_1(1420)$, then it could also be contributed through other partial-wave analysis, for example, $f_1(1285) \to K_S^0 K^+ \pi^-$~\cite{He:2021exv}. With the theoretical predictions and experimental measurements of ${\cal B}(B_s^0 \to J/\psi f_1(1285))$  and ${\cal B}(f_1(1285) \to K\bar K \pi) = 0.090 \pm  0.004$ ~\cite{ParticleDataGroup:2022pth}, the magnitude of  ${\cal B}(B_s^0 \to J/\psi f_1(1285)(\to K_S^0 K^\pm \pi^\mp))$ is about $10^{-6} \sim 10^{-5}$, which is measurable in the on-going LHCb and Belle-II experiments. Meanwhile, combining our results and the available experimental data ${\cal B}(f_{1}(1285) \to \eta \pi^{+} \pi^{-}) =0.35_{-0.15}^{+0.15}$ \cite{ParticleDataGroup:2022pth} and ${\cal B}(f_1(1420) \to K_S^0 K^\pm \pi^\mp) \approx 0.64_{-0.06}^{+0.06}$~\cite{Jiang:2020eml}, we then estimate ${\cal B}(B^0 \to J/\psi f_1(1285)(\to \eta \pi^+ \pi^-))$ and ${\cal B}(B^0 \to J/\psi f_1(1420)(\to K_S^0 K^+ \pi^-))$ as
\beq
{\cal B}(B_d^0 \to J/\psi (\eta \pi^+\pi^-)_{f_1(1285)})&\equiv&
{\cal B}(B_d^0 \to J/\psi f_1(1285))
\non &&
\cdot {\cal B}(f_1(1285) \to \eta \pi^+ \pi^-)
\approx
\biggl\{\begin{array}{ll}
(1.07^{+0.84}_{-0.70})
 \times 10^{-5}
 \\
(0.96^{+0.75}_{-0.63}) \times 10^{-5}
 \label{eq:br-f12d-epp}
\end{array}\;,
\\
{\cal B}(B_s^0 \to J/\psi (\eta \pi^+\pi^-)_{f_1(1285)})&\equiv&
{\cal B}(B_s^0 \to J/\psi f_1(1285))
\non &&
\cdot {\cal B}(f_1(1285) \to \eta \pi^+ \pi^-)
\approx
\biggl\{\begin{array}{ll}
(0.31^{+0.21}_{-0.18})
 \times 10^{-4}
 \\
(0.77^{+0.54}_{-0.47})  \times 10^{-4}
 \label{eq:br-f12s-epp}
\end{array}\;,
\\
{\cal B}(B_d^0 \to J/\psi (K_S^0 K^+ \pi^-)_{f_1(1420)})&\equiv&
{\cal B}(B_d^0 \to J/\psi f_1(1420))
\non &&
\cdot {\cal B}(f_1(1420) \to K_S^0 K^+ \pi^-)
\approx
\biggl\{\begin{array}{ll}
(1.27^{+0.88}_{-0.68})  \times 10^{-6}
 \\
(3.15^{+2.24}_{-1.72})  \times 10^{-6}
 \label{eq:br-f14d-kkp}
\end{array}\;,
\\
{\cal B}(B_s^0 \to J/\psi (K_S^0 K^+ \pi^-)_{f_1(1420)})&\equiv&
{\cal B}(B_s^0 \to J/\psi f_1(1420))
\non &&
\cdot {\cal B}(f_1(1420) \to K_S^0 K^+ \pi^-)
\approx
\biggl\{\begin{array}{ll}
(0.73^{+0.36}_{-0.28})  \times 10^{-3}
 \\
(0.65^{+0.33}_{-0.25}) \times 10^{-3}
 \label{eq:br-f14s-kkp}
\end{array}\;,
\eeq
where all uncertainties have been added in quadrature. In above results, the upper and the lower entries correspond to the results obtained with $|\varphi^{\rm Theo}| \sim 15^\circ$ and $|\varphi^{\rm Exp}| \sim 24^\circ$, respectively (In the following context, the similar presentation will be adopted implicitly, unless otherwise stated.). The results of Eqs.~(\ref{eq:br-f12d-epp})-(\ref{eq:br-f14s-kkp}) are expected to be tested at LHCb and Belle-II experiments in the near future, and they could further provide more additional constraints on $|\varphi|$ and rich information for understanding the nature of $f_1$.

In PDG \cite{ParticleDataGroup:2022pth}, the branching fractions of $B_d^0 \to J/\psi \rho^0$ and $B_s^0 \to J/\psi \phi$ have been well measured as
\beq
{\cal B}(B_{d}^0 \to J/\psi \rho^0) &=&
(2.55^{+0.18}_{-0.16})\times 10^{-5}\;,
\qquad
{\cal B}(B_{s}^0 \to J/\psi\phi) =
(1.04^{+0.04}_{-0.04})\times 10^{-3}\;,
\eeq
These branching fractions with high precision can be used for normalizations in studying $B \to J/\psi f_1$ decays. Theoretically, both branching fractions have been calculated in PQCD at NLO \cite{Liu:2013nea} are updated, and the results are given as
\beq
{\cal B}(B_{d}^0 \to J/\psi \rho^0) &=&
(2.98^{+0.81}_{-0.69})\times 10^{-5}\;,
\qquad
{\cal B}(B_{s}^0 \to J/\psi\phi) =
(1.07^{+0.33}_{-0.29})\times 10^{-3}\;,
\eeq
where the uncertainties arising from different sources have been added in quadrature. Considering  ${\cal B}(\phi \to K^+ K^-) = 0.492 \pm 0.005$ and ${\cal B}(\rho^0 \to \pi^+ \pi^-) \sim 100\%$~\cite{ParticleDataGroup:2022pth}, we then obtain the branching fractions of $B_s^0 \to J/\psi \phi (\to K^+ K^-)$ and $B_d^0 \to J/\psi \rho^0 (\to \pi^+ \pi^-)$ decays as follows,
\beq
{\cal B}(B_{s}^0 \to J/\psi (K^{+} K^{-})_{\phi}) &\equiv&
{\cal B}(B_{s}^0 \to J/\psi \phi) \cdot {\cal B}(\phi \to K^{+}K^{-})
\approx
(0.53^{+0.16}_{-0.14}) \times 10^{-3}\;,
\\
{\cal B}(B_{d}^0 \to J/\psi (\pi^{+} \pi^{-})_{\rho^0}) &\equiv&
{\cal B}(B_{d}^0 \to J/\psi \rho^0) \cdot {\cal B}(\rho^0 \to \pi^{+} \pi^{-})
\approx
(2.98^{+0.81}_{-0.69}) \times 10^{-5}\;.
\eeq
Then, four ratios could be defined as follows,
\beq
R_{s}^{J/\psi}[f_{1}(1420)/\phi]
&\equiv& \frac{{\cal B}(B_{s}^{0} \to J/\psi f_{1}(1420)) }
{{\cal B}(B_{s}^0 \to J/\psi \phi) }
\approx
\biggl\{\begin{array}{cc}
1.07^{+0.34}_{-0.25}
\\
0.95^{+0.31}_{-0.23}
\end{array}\;,
\\
R_{d}^{J/\psi}[f_{1}(1285)/\rho^0]&\equiv&
\frac{{\cal B}(B_{d}^{0} \to J/\psi f_{1}(1285))}
{{\cal B}(B_{d}^0 \to J/\psi \rho^0)}
\approx
\biggl\{\begin{array}{cc}
1.02^{+0.55}_{-0.42}
\\
0.92^{+0.49}_{-0.37}
\end{array}\;,
\eeq
and
\beq
R_{\pi}^{J/\psi}&\equiv&
\frac{{\cal B}(B_s^0 \to J/\psi (K_S^0 K^+ \pi^-)_{f_1(1420)})}
{{\cal B}(B_{s}^0 \to J/\psi (K^{+} K^{-})_{\phi}) }
\approx
\biggl\{\begin{array}{cc}
1.38^{+0.48}_{-0.34}
\\
1.23^{+0.41}_{-0.30}
\end{array}\;,
\\
R_\eta^{J/\psi} &\equiv&
\frac{{\cal B}(B_d^0 \to J/\psi (\eta \pi^+\pi^-)_{f_1(1285)})}
{{\cal B}(B_{d}^0 \to J/\psi (\pi^{+} \pi^{-})_{\rho^0})}
\approx
\biggl\{\begin{array}{cc}
0.36^{+0.24}_{-0.21}
\\
0.32^{+0.22}_{-0.19}
\end{array}\;.
\eeq
All the above results could be tested in LHCb and Belle-II experiments.  As far as the central values are concerned, because the behaviors of $f_1$ meson are very similar to those of $\rho/\phi$ \cite{Yang:2007zt,Cheng:2007mx,Yang:2008xw,Cheng:2008gxa}, the results indicate that  $f_n(f_s)$ state should predominately govern the $f_1(1285)(f_1(1420))$ meson, that is, the small $|\varphi|$ is preferred.

In Ref.~\cite{Aaij:2013rja}, after neglecting the contributions from penguin operators and assuming the SU(3) flavor symmetry, LHCb estimated the angle $\varphi$ with the relation
\begin{equation}
\label{eq:tansqphi}
\tan^2\varphi\approx\frac{1}{2}\frac{{\cal B}(B_{s}^{0} \to J/\psi f_{1}(1285))}{{\cal B}(B_{d}^{0} \to J/\psi f_{1}(1285))}
\frac{\tau_0}{\tau_s}\frac{|V_{cd}|^2}{|V_{cs}|^2}
\frac{\Phi(m_{B_d},m_{J/\psi},m_{f_{1}(1285)})}
     {\Phi(m_{B_s},m_{J/\psi},m_{f_{1}(1285)})},
\end{equation}
where the phase space factor is given as $\Phi(a,b,c)=\left[(a^2-(b+c)^2)(a^2-(b-c)^2)\right]^{\frac{3}{2}}$. In fact, the corrections from penguin operators are about $15\%$ \cite{Li:2006vq}, and SU(3) asymmetry is particularly estimated to be about $30\%$. Without above uncertainties, we define two ratios and evaluate them as
\beq
R_{d}^{J/\psi}&\equiv&
\frac{{\cal B}(B_{d}^{0} \to J/\psi f_{1}(1285))}
{{\cal B}(B_{d}^{0} \to J/\psi f_{1}(1420))}
= \frac{\Phi(m_{B_d},m_{J/\psi},m_{f_{1}(1285)})}{\Phi(m_{B_d},m_{J/\psi},m_{f_{1}(1420)})}\cdot \cot^2\varphi
\approx
\biggl\{\begin{array}{rr}
15.33^{+0.15}_{-0.14}
\\
5.55^{+0.06}_{-0.04}
\end{array}\;,
\label{eq:brr-d12-d14-jp}
\\
R_{s}^{J/\psi}&\equiv&
\frac{{\cal B}(B_{s}^{0} \to J/\psi f_{1}(1420))}
{{\cal B}(B_{s}^{0} \to J/\psi f_{1}(1285))}
= \frac{\Phi(m_{B_s},m_{J/\psi},m_{f_{1}(1420)})}{\Phi(m_{B_s},m_{J/\psi},m_{f_{1}(1285)})}\cdot \cot^2\varphi
\approx
\biggl\{\begin{array}{rr}
12.81^{+0.17}_{-0.18}
\\
4.62^{+0.08}_{-0.03}
\end{array}\;.
\label{eq:brr-s14-s12-jp}
\eeq
Such two ratios with small uncertainties could also be used to determine $|\varphi|$ in future, if $f_{1}(1285)$ and $f_{1}(1420)$ are believed to be two-quark states.

Now we turn to other observables such as the polarization fractions and the direct CP asymmetries in $B^0 \to J/\psi f_1$ decays. With the helicity amplitudes shown in Eqs.~(\ref{eq:DecAmp-psif12-d})-(\ref{eq:DecAmp-psif14-s}), a set of transversity amplitudes, namely, the longitudinal one ${\cal A}_L$, the parallel one ${\cal A}_{\parallel}$, and the perpendicular one ${\cal A}_{\perp}$, can be defined respectively as follows,
\beq
{\cal A}_{L} &=& {\cal A}_{L}\;,
\qquad
{\cal A}_{\parallel}=\sqrt{2} {\cal A}_{N}\;,
\qquad
{\cal A}_{\perp}=r_{2}r_{3}\sqrt{2(\kappa^{2}-1)}{\cal A}_{T}\;,
\label{eq:Tamp}
\eeq
with the ratio $\kappa=P_{2} \cdot P_{3}/(m_{J/\psi}m_{f_1})$. The polarization fractions $f_{L}$ and $f_T$ in $B^0 \to J/\psi f_1$ decays  can be defined as follows~\cite{BaBar:2007bpi},
\beq
f_{L}&\equiv& \frac{|{\cal A}_{L}|^2}{|{\cal A}_L|^2+|{\cal A}_{||}|^2+|{\cal A}_{\perp}|^2}\;, \qquad
f_T \equiv \frac{|{\cal A}_{||}|^2+ |{\cal A}_\perp|^2}{|{\cal A}_L|^2+|{\cal A}_{||}|^2+|{\cal A}_{\perp}|^2} = f_\parallel + f_\perp \;,
\label{eq:pf}
\eeq
which satisfy the relation of $f_{L}+f_{T}=1$. The relative phases $\phi_{\parallel}$ and $\phi_\perp$ (in units of rad) are thus obtained as follows,
\beq
\phi_{\parallel}&=&\arg\frac{{\cal A}_{\parallel}}{{\cal A}_{L}}\;,
\qquad
\phi_{\perp}=\arg\frac{{\cal A}_{\perp}}{{\cal A}_{L}}\;.
\eeq
In PQCD, these observables of  $B^0 \to J/\psi f_1$ decays are calculated, and the results are given as
\beq
f_{L}(B_{d}^{0} \to J/\psi f_{1}(1285)) &=&
(44.6^{+13.9}_{-10.4}) \%\;,
\qquad
f_{L}(B_{d}^{0} \to J/\psi f_{1}(1420)) =
(45.8^{+13.8}_{-10.4}) \%\;,
\nonumber
\\
f_{L}(B_{s}^{0} \to J/\psi f_{1}(1285)) &=&
(45.1^{+14.1}_{-10.7}) \%\;,
\qquad
f_{L}(B_{s}^{0} \to J/\psi f_{1}(1420)) =
(46.2^{+14.0}_{-10.9}) \%\;,
\label{eq:pf-s-psif14}
\eeq
and
\beq
\phi_{\parallel}(B_d^0 \to J/\psi f_1)&=&2.11_{-0.05}^{+0.05}\;,
\qquad
\phi_{\perp}(B_d^0 \to J/\psi f_1)=1.99_{-0.07}^{+0.05}\;, \nonumber
\\
\phi_{\parallel}(B_s^0 \to J/\psi f_1)&=&2.15_{-0.06}^{+0.09}\;,
\qquad
\phi_{\perp}(B_s^0 \to J/\psi f_1)=2.01_{-0.06}^{+0.06}\;,
\eeq
in which various errors have been added in quadrature. Though these quantities are  the ratios of (squared) decay amplitudes, the nonperturbative parameters, especially the Gegenbauer moment $a_1^\perp$ in the distribution amplitudes of $f_n$ and $f_s$ and the charm quark mass $m_c$, take large theoretical uncertainties. We also note that the longitudinal polarization fractions and transverse ones are almost equal, due to the helicity flip. Such phenomenon has been confirmed in $B_s\to J/\psi \phi$ and $B_d\to J/\psi K^*$ decays \cite{Liu:2013nea, LHCb:2013vga, LHCb:2011aa}. All these observables will also be tested in experiments.

The direct CP asymmetry $A_{\rm CP}^{\rm dir}$ of $B^0 \to J/\psi f_1$ decays is defined as
\beq
A_{\rm CP}^{\rm dir}
&\equiv&  \frac{|\overline{\cal A}(\overline{B}^0\to \overline f)|^2 - |{\cal A}(B^0\to f)|^2}
{|\overline{\cal A}(\overline{B}^0\to \overline f)|^2+|{\cal A}(B^0\to f)|^2}\;,
\label{eq:acp1}
\eeq
where ${\cal A}$ stand for the decay amplitudes of $B^0 \to J/\psi f_{1}$, while $\overline{\cal A}$ describe the corresponding charge conjugation ones. With the obtained amplitudes, we calculate the direct CP asymmetries and present the results as
\beq
A_{\rm CP}^{\rm dir}(B_{d}^{0} \to J/\psi f_{1})
&=& A_{\rm CP}^{\rm dir}(B_d^0 \to J/\psi f_n)
=
(-5.86^{+4.54}_{-5.52})
\times 10^{-3}\;,
\\
A_{\rm CP}^{\rm dir}(B_{s}^{0} \to J/\psi f_{1})
&=&
A_{\rm CP}^{\rm dir}(B_{s}^{0} \to J/\psi f_s)
=
(2.24^{+2.41}_{-1.97})
\times 10^{-4}\;.
\eeq
Meanwhile, the direct CP asymmetries in each polarization can also be studied as \cite{Beneke:2006hg}
\beq
A_{\rm CP}^{\rm dir,\alpha}&=&\frac{\bar{f}_{\alpha}-f_{\alpha}}{\bar{f}_{\alpha}+f_{\alpha}}
\quad (\alpha=L, \parallel, \perp)\;,
\eeq
where $\bar f_\alpha$ is the polarization fraction for the corresponding $\bar B$ decays in Eq.~(\ref{eq:pf}). In this work, the direct CP asymmetries for $B^0 \to J/\psi f_{1}$ decays at each polarization  are collected as follows,
\begin{itemize}
\item[{$\bullet$}]{for $B_{d}^{0} \to J/\psi f_{1}$ modes (in units of $10^{-3}$),}
\beq
A_{\rm CP}^{\rm dir,L}&=& -5.05_{-7.83}^{+5.82}
,
 \qquad
A_{\rm CP}^{\rm dir,\parallel} = -6.38_{-4.46}^{+4.02}
,
 \qquad
A_{\rm CP}^{\rm dir,\perp} = -6.78_{-4.07}^{+3.55}
,
\eeq

\item[{$\bullet$}]{for $B_{s}^{0} \to J/\psi f_{1}$ modes (in units of $10^{-4}$),}
\beq
A_{\rm CP}^{\rm dir,L} &=& 1.69_{-2.17}^{+2.86}
,
   \qquad
A_{\rm CP}^{\rm dir,\parallel} = 2.58_{-2.03}^{+2.35}
,
   \qquad
A_{\rm CP}^{\rm dir,\perp} = 2.89_{-1.72}^{+2.05}
.
\eeq
\end{itemize}
All uncertainties from various parameters in above results have been added in quadrature. It is obvious that the direct CP violations for $B_d^0 \to J/\psi f_1$ decays are a few percent within errors, however, which are still too small to be detected in the running LHCb and Belle-II experiments.

\subsection{ 
\boldmath{$B \to \eta_c f_1$}
}
\label{ssec:etacf1}

Different from the decays of $B^0 \to J/\psi f_1$, $B^0 \to \eta_c f_1$ decays have only the longitudinal contributions because of the conservation of the angular momentum. By employing the decay amplitudes as presented in Eqs.~(\ref{eq:DecAmp-etacf12-d})-(\ref{eq:DecAmp-etacf14-s}), the corresponding branching ratio can be expressed as follows,
\beq
{\cal B}(B^0 \to \eta_c f_1) &=& \tau_{B} \cdot
\frac{G_F^2}{32\pi m_B} (1-r_{\eta_c}^2) |{\cal A}(B^0 \to \eta_c f_1)|^2\;,
\eeq
with $r_{\eta_c} = m_{\eta_c}/m_{B^0}$. Similar to ${\cal B}(B^0 \to J/\psi f_1)$, the theoretical predictions of ${\cal B}(B^0 \to \eta_c f_1)$ at both $|\varphi^{\rm Theo}|$ and $|\varphi^{\rm Exp}|$ in PQCD approach are presented in Table.~\ref{tab:br-etacf1}, where the sources of errors are also similar to those in $B^0 \to J/\psi f_1$ modes but with $f_{\eta_c} = 0.42 \pm 0.05$~GeV. Within the large theoretical errors, the $B^0 \to \eta_c f_1$ decay rates at $|\varphi^{\rm Theo}| \sim 15^\circ$ are generally consistent with those at $|\varphi^{\rm Exp}| \sim 24^\circ$. These decays have not been measured yet currently, and are expected to be tested in the near future.
\begin{table}[hbt]
\caption{ The CP-averaged branching ratios $B^0 \to \eta_c f_1$ in PQCD approach}
\label{tab:br-etacf1}
 \begin{center}\vspace{-0.3cm}{
\begin{tabular}[t]{c|c|c}
\hline  \hline
  Decay Modes   &  $|\varphi^{\rm Theo}| \sim 15^\circ$
&  $|\varphi^{\rm Exp}| \sim 24^\circ$
\\
\hline \hline
  ${\cal B}(B_{d}^{0} \to \eta_c f_{1}(1285))$
&
$(7.64^{+2.42+5.95+0.58+0.10}_{-1.78-4.49-0.66-0.11})\times 10^{-6}$
 &
$(6.83^{+2.16+5.34+0.52+0.28}_{-1.59-4.02-0.58-0.35})\times 10^{-6}$
 \\
 \hline
  ${\cal B}(B_{d}^{0} \to \eta_c f_{1}(1420))$
&
$(5.16^{+1.64+4.04+0.40+1.06}_{-1.20-3.03-0.44-0.96})\times 10^{-7}$
 &
$(1.28^{+0.40+0.99+0.09+0.33}_{-0.30-0.76-0.11-0.26})\times 10^{-6}$
 \\
 \hline
  ${\cal B}(B_{s}^{0} \to \eta_c f_{1}(1285))$
&
$(2.15^{+0.81+1.42+0.18+0.44}_{-0.58-1.08-0.18-0.40})\times 10^{-5}$
&
$(5.31^{+2.00+3.45+0.45+1.39}_{-1.43-2.68-0.46-1.08})\times 10^{-5}$
 \\
 \hline
  ${\cal B}(B_{s}^{0} \to \eta_c f_{1}(1420))$
&
$(2.84^{+1.07+1.88+0.24+0.03}_{-0.77-1.43-0.25-0.05})\times 10^{-4}$
&
$(2.54^{+0.96+1.68+0.22+0.10}_{-0.69-1.28-0.22-0.13})\times 10^{-4}$
\\
 \hline \hline
\end{tabular}}
\end{center}
\end{table}

Comparing $B^0 \to J/\psi f_1$  and $B^0 \to \eta_c f_1$ decays, we have
\beq
\frac{{\cal B}(B_{d}^{0} \to J/\psi f_{1}(1285))}
{{\cal B}(B_{d}^{0}\to \eta_{c} f_{1}(1285))}
\approx
4.00^{+1.29}_{-0.85}\;,
&\qquad&
\frac{{\cal B}(B_{s}^{0} \to J/\psi f_{1}(1285))}
{{\cal B}(B_{s}^{0}
\to \eta_{c} f_{1}(1285))}
\approx
4.14^{+1.48}_{-0.90} \;,
\nonumber
\\
\frac{{\cal B}(B_{d}^{0} \to J/\psi f_{1}(1420))}
{{\cal B}(B_{d}^{0}
\to \eta_{c} f_{1}(1420))}
\approx
3.86^{+1.27}_{-0.82} \;,
&\qquad&
\frac{{\cal B}(B_{s}^{0} \to J/\psi f_{1}(1420))}
{{\cal B}(B_{s}^{0}
\to \eta_{c} f_{1}(1420))}
\approx
4.01^{+1.51}_{-0.90} \;.
\eeq
It is noticeable that these ratios are all about 4, which attributes to the large contributions from the transverse polarization. The measurements of such kinds of ratios in future could help us to test the polarization mechanism.

Analogous to the decays $B^0 \to J/\psi f_1$, we also define four ratios in $B^0 \to \eta_c f_1$ modes as follows,
\begin{eqnarray}
R_{d}^{\eta_c} &\equiv&
\frac{{\cal B}(B_{d}^{0} \to \eta_{c} f_{1}(1285))}
     {{\cal B}(B_{d}^{0} \to \eta_{c} f_{1}(1420))}
=
\frac{\Phi(m_{B_d},m_{\eta_c},m_{f_{1}(1285)})}
     {\Phi(m_{B_d},m_{\eta_c},m_{f_{1}(1420)})}
\cdot \cot^2\varphi \;, \label{eq:brr-d12-d14-et}\\
R_{s}^{\eta_c} &\equiv&
\frac{{\cal B}(B_{s}^{0} \to \eta_{c} f_{1}(1420))}
     {{\cal B}(B_{s}^{0} \to \eta_{c} f_{1}(1285))}
=
\frac{\Phi(m_{B_s},m_{\eta_c},m_{f_{1}(1420)})}
     {\Phi(m_{B_s},m_{\eta_c},m_{f_{1}(1285)})}
\cdot \cot^2\varphi \;, \label{eq:brr-s14-s12-et}\\
R_{sd}^{\eta_c}[f_1(1285)]&\equiv&
\frac{{\cal B}(B_{s}^{0} \to \eta_{c} f_{1}(1285))}
     {{\cal B}(B_{d}^{0} \to \eta_{c} f_{1}(1285))}\nonumber \\
&=&
\frac{\tau_{B_s^0}}{\tau_{B_d^0}}
\cdot \frac{\Phi(m_{B_s},m_{\eta_c},m_{f_{1}(1285)})}
           {\Phi(m_{B_d},m_{\eta_c},m_{f_{1}(1285)})}
\cdot \frac{|{\cal A}(B_s^0 \to \eta_c f_s)|^2}
           {|{\cal A}(B_d^0 \to \eta_c f_n)|^2}
\cdot \tan^2\varphi \\
R_{sd}^{\eta_c}[f_1(1420)]&\equiv&
\frac{{\cal B}(B_{s}^{0} \to \eta_{c} f_{1}(1420))}
     {{\cal B}(B_{d}^{0} \to \eta_{c} f_{1}(1420))}\nonumber \\
&=&
\frac{\tau_{B_s^0}}{\tau_{B_d^0}}
\cdot \frac{\Phi(m_{B_s},m_{\eta_c},m_{f_{1}(1420)})}
           {\Phi(m_{B_d},m_{\eta_c},m_{f_{1}(1420)})}
\cdot \frac{|{\cal A}(B_s^0 \to \eta_c f_s)|^2}
           {|{\cal A}(B_d^0 \to \eta_c f_n)|^2}
\cdot \cot^2\varphi \;,
\end{eqnarray}
which can be used to constrain the absolute value of the mixing angle $\varphi$. In PQCD, these above ratios could be calculated from Table.~\ref{tab:br-etacf1}, and the results are given as
\begin{eqnarray}
R_{d}^{\eta_c}
\approx
\biggl\{\begin{array}{cc}
14.81^{+3.12}_{-2.36}
\\
5.34^{+1.02}_{-0.92}
\end{array}\;,
&\quad&
R_{s}^{\eta_c}
\approx
\biggl\{\begin{array}{cc}
13.21^{+2.73}_{-2.13}
\\
4.78^{+0.91}_{-0.84}
\end{array}\;,\nonumber
\\
R_{sd}^{\eta_c}[f_1(1285)]\approx
\biggl\{\begin{array}{cc}
2.81^{+1.32}_{-0.94}
\\
7.77^{+3.76}_{-2.56}
\end{array}\;,
&\quad&
R_{sd}^{\eta_c}[f_1(1420)]
\approx
\biggl\{\begin{array}{cc}
(5.50^{+2.64}_{-1.83}) \times 10^{2},
\\
(1.98^{+0.99}_{-0.72}) \times 10^{2},
\end{array}
\eeq
where all uncertainties are added in quadrature.

Since the branching fractions of $B_{d,s}^0 \to J/\psi f_1(1285)$ decays were measured with ${\cal B}(f_1(1285) \to 2\pi^+ 2\pi^-) =0.109_{-0.006}^{+0.006}$,  we also propose the analogous measurements on $B_{d,s}^0 \to \eta_c f_1(1285)(\to 2\pi^+ 2\pi^-)$, whose branching fractions are estimated to be
\beq
{\cal B}(B_d^0 \to \eta_{c} (2\pi^+2\pi^-)_{f_1(1285)})
&\equiv&
{\cal B}(B_d^0 \to \eta_{c} f_1(1285))
\non &&
\cdot {\cal B}(f_1(1285) \to 2\pi^+2\pi^-)
\approx
\biggl\{\begin{array}{cc}
( 0.84^{+0.71}_{-0.54})
\times 10^{-6}
\\
( 0.75^{+0.64}_{-0.48})
\times 10^{-6}
\end{array}\;,
\\
{\cal B}(B_s^0 \to \eta_{c} (2\pi^+2\pi^-)_{f_1(1285)})
&\equiv&
{\cal B}(B_s^0 \to \eta_{c} f_1(1285))
\non &&
\cdot {\cal B}(f_1(1285) \to 2\pi^+2\pi^-)
\approx
\biggl\{\begin{array}{cc}
( 0.24^{+0.19}_{-0.14})
\times 10^{-5}
\\
( 0.58^{+0.49}_{-0.36})
\times 10^{-5}
\end{array}\;.
\eeq
Meanwhile, with the large widths of $f_1(1285) \to \eta \pi^+ \pi^-$ and $f_1(1420) \to K_S^0 K^+ \pi^-$, the branching fractions of $B^0 \to \eta_c f_1(1285)(\to \eta \pi^+ \pi^-)$ and $B^0 \to \eta_c f_1(1420)(\to K_s^0 K^+ \pi^-)$ decays are also calculated to be
\beq
{\cal B}(B_d^0 \to \eta_{c} (\eta \pi^+\pi^-)_{f_1(1285)})
&\equiv&
{\cal B}(B_d^0 \to \eta_{c} f_1(1285))
\non &&
\cdot {\cal B}(f_1(1285) \to \eta \pi^+ \pi^-)
\approx
\biggl\{\begin{array}{cc}
( 2.67^{+2.54}_{-2.06})
\times 10^{-6}
\\
( 2.39^{+2.27}_{-1.84})
\times 10^{-6}
\end{array}\;,
\\
{\cal B}(B_s^0 \to \eta_{c} (\eta \pi^+\pi^-)_{f_1(1285)})
&\equiv&
{\cal B}(B_s^0 \to \eta_{c} f_1(1285))
\non &&
\cdot
{\cal B}(f_1(1285) \to \eta \pi^+ \pi^-)
\approx
\biggl\{\begin{array}{cc}
( 0.75^{+0.68}_{-0.56})
\times 10^{-5}
\\
( 1.86^{+1.71}_{-1.37})
\times 10^{-5}
\end{array}\;,
\\
{\cal B}(B_d^0 \to \eta_{c} (K_S^0 K^+ \pi^-)_{f_1(1420)})
&\equiv&
{\cal B}(B_d^0 \to \eta_{c} f_1(1420))
\non &&
\cdot
{\cal B}(f_1(1420) \to K_S^0 K^+ \pi^-)
\approx
\biggl\{\begin{array}{cc}
( 3.30^{+2.90}_{-2.22})
\times 10^{-7}
\\
( 0.82^{+0.72}_{-0.56})
\times 10^{-6}
\end{array}\;,
\\
{\cal B}(B_s^0 \to \eta_{c} (K_S^0 K^+ \pi^-)_{f_1(1420)})
&\equiv&
{\cal B}(B_s^0 \to \eta_{c} f_1(1420))
\non &&
\cdot
{\cal B}(f_1(1420) \to K_S^0 K^+ \pi^-)
\approx
\biggl\{\begin{array}{cc}
( 1.82^{+1.40}_{-1.07})
\times 10^{-4}
\\
( 1.63^{+1.26}_{-0.96})
\times 10^{-4}
\end{array}\;.
\eeq
It is obvious that the decays $B_s^0 \to \eta_c f_1(1285)(\to 2\pi^+ 2\pi^-)$, $B^0 \to \eta_c f_1(1285)(\to \eta \pi^+ \pi^-)$ and $B_s^0 \to \eta_c f_1(1420)(\to K_S^0 K^+ \pi^-)$ with large decay rates could be observed in the running LHCb and Belle-II experiments. Once these predictions would be confirmed experimentally, then the mixing angle $\varphi$ between the $f_1(1285)$ and $f_1(1420)$ mixing could receive more constraints, though they still suffer from large theoretical uncertainties induced by the hadronic parameters in the $f_n$ and $f_s$ light-cone distribution amplitudes.

In measuring the branching fractions of $B_{d,s}^0 \to \eta_c f_1$ decays, we can also select the modes $B_d^0 \to \eta_c \rho^0$ and $B_s^0 \to \eta_c \phi$ for normalization. Theoretically, the ${\cal B}(B_d^0 \to \eta_c \rho^0)$ and ${\cal B}(B_s^0 \to \eta_c \phi)$ in Ref.~\cite{Xiao:2019mpm} are updated, and the results are presented as follows,
\beq
{\cal B}(B_{d}^0 \to \eta_{c} \rho^0)=
 (7.92^{+3.67}_{-3.04}) \times 10^{-6}\;,
 \qquad
{\cal B}(B_{s}^0 \to \eta_{c} \phi) =
(3.57^{+1.81}_{-1.41} )\times 10^{-4}\;,
\eeq
which are slightly smaller than but consistent with previous predictions \cite{Xiao:2019mpm}. The updated ${\cal B}(B_s^0 \to \eta_c \phi)$ still agrees well with data ${\cal B}(B_s^0 \to \eta_c \phi)=(5.0\pm 0.9 )\times 10^{-4}$  \cite{ParticleDataGroup:2022pth} within errors, but the branching fraction of $B_{d}^0 \to \eta_{c} \rho^0$ have not been measured so far. Within  $\rho^0 \to \pi^+ \pi^-$ and $\phi \to K^+ K^-$, we then obtain the following results as,
\beq
{\cal B}(B_{d}^0 \to \eta_{c} (\pi^{+} \pi^{-})_{\rho^0}) &\equiv&
{\cal B}(B_{d}^0 \to \eta_{c} \rho^0) \cdot {\cal B}(\rho^0 \to \pi^{+} \pi^{-})
\approx
(7.92^{+3.67}_{-3.04})
\times 10^{-6}\;,
\\
{\cal B}(B_{s}^0 \to \eta_{c} (K^{+} K^{-})_{\phi}) &\equiv&
{\cal B}(B_{s}^0 \to \eta_{c} \phi) \cdot {\cal B}(\phi \to K^{+}K^{-})
\approx
(1.76^{+0.89}_{-0.69})
\times 10^{-4}\;.
\eeq
In order to search for $B^0 \to \eta_c f_1$ decays, we also calculate the following ratios
\beq
R_d^{\eta_c}[f_{1}(1285)/\rho^0]&\equiv&
\frac{{\cal B}(B_{d}^{0} \to \eta_{c} f_{1}(1285))}
{{\cal B}(B_{d}^0 \to \eta_{c} \rho^0)}
\approx
\biggl\{\begin{array}{cc}
0.96^{+0.39}_{-0.31}
\\
0.86^{+0.35}_{-0.27}
\end{array}\;,
\\
R_s^{\eta_c}[f_{1}(1420)/\phi]&\equiv&
\frac{{\cal B}(B_{s}^{0} \to \eta_{c} f_{1}(1420))}
{{\cal B}(B_{s}^0 \to \eta_{c} \phi)}
\approx
\biggl\{\begin{array}{cc}
0.80^{+0.08}_{-0.08}
\\
0.71^{+0.06}_{-0.06}
\end{array}\;,
\eeq
and
\beq
R^{\eta_c}_{\pi} &\equiv&
\frac{{\cal B}(B_s^0 \to \eta_{c} (K_S^0 K^+ \pi^-)_{f_1(1420)})}
{{\cal B}(B_{s}^0 \to \eta_{c} (K^{+} K^{-})_{\phi})}
\approx
\biggl\{\begin{array}{cc}
1.03^{+0.14}_{-0.13}
\\
0.93^{+0.11}_{-0.12}
\end{array} \;,
\\
R^{\eta_c}_{\eta}&\equiv&
\frac{{\cal B}(B_d^0 \to \eta_{c} (\eta \pi^+\pi^-)_{f_1(1285)})}
{{\cal B}(B_{d}^0 \to \eta_{c} (\pi^{+} \pi^{-})_{\rho^0})}
\approx
\biggl\{\begin{array}{cc}
0.34^{+0.20}_{-0.18}
\\
0.30^{+0.18}_{-0.16}
\end{array} \;,
\eeq
These ratios are also expected to be examined in future.

Lastly, we shall study the direct CP asymmetries of $B^0 \to J/\psi f_1$ decays. Based on the definition in Eq.~(\ref{eq:acp1}), the numerical results can be given as
\beq
A_{\rm CP}^{\rm dir}(B_{d}^{0} \to \eta_{c} f_{1})
&=&
A_{\rm CP}^{\rm dir}(B_{d}^{0} \to \eta_{c} f_n)
=
(7.53^{+1.07}_{-0.98})
\times 10^{-2}\;,
\\
A_{\rm CP}^{\rm dir}(B_{s}^{0} \to \eta_{c} f_{1})
&=&
 A_{\rm CP}^{\rm dir}(B_{s}^{0} \to \eta_{c} f_s)
=
(-4.33^{+0.60}_{-0.66})
\times 10^{-3}\;.
\eeq
It is shown that the direct CP asymmetries of $B \to \eta_c f_1$  are as large as a few percent, and are an order of magnitude larger than those of $B^0 \to J/\psi f_1$ decays. The measurements of such asymmetries in the running LHCb and Belle-II experiments are helpful for understanding the nature of $f_1$ states and testing the PQCD approach.

%
%
\section{
Summary}
\label{sect:4}
We have  made reexaminations on $B^0 \to J/\psi f_1$ and the first time studies on $B^0 \to \eta_c f_1$ decays in the framework of PQCD, where $f_1=f_1(1280)$ and $f_1(1420)$ are viewed as the mixtures of $f_n$ and $f_s$ with mixing angle $\varphi$. In this work, the contributions of vertex corrections, nonfactorizable diagrams and penguin operators are all included. It is found that the branching fractions of these decays are large enough to be measured in the running LHCb and Belle-II experiments. We also note that the $B^0 \to \eta_c f_1(1285)$ and $B_s^0 \to (J/\psi, \eta_c) f_1(1420)$ decays could be analyzed within the secondary decay chains $f_1(1285) \to 2\pi^+ 2\pi^-$, $f_1(1285) \to \eta \pi^+ \pi^-$ and $f_1(1420) \to K_S^0 K^+ \pi^-$. In addition, we proposed several ratios that could be used to constrain the mixing angle $\varphi$, but its sign cannot be determined in these decays. We also studied the  direct CP asymmetries in these decays, and results indicate the large penguin pollution in the $B_d^0 \to (J/\psi, \eta_c) f_1$ decays. It should be emphasized that there are large theoretical uncertainties arising from the nonperturbative parameters, especially from the distribution amplitudes of axial-vector mesons and charmonium states, and more precise parameters from nonperturbative QCD approaches are needed. The comparisons between our results and future experimental data would help us to understand the nature of $f_1$ states and to test the PQCD approach.

%
%

\begin{acknowledgments}
D.Y. thanks Z.J. for his helpful discussions.
This work is supported by the National Natural Science
Foundation of China under Grants nos.~11875033, 11975195, and 11775117,
by the Qing Lan Project of Jiangsu Province under Grant No.~9212218405,
by the Natural Science Foundation of Shandong province under the Grants nos.
~ZR2019JQ04, ZR2022MA035 and ZR2022ZD26, by the Project of Shandong Province Higher Educational Science and Technology Program under Grant no.~2019KJJ007,
and by the Research Fund of Jiangsu Normal University under Grant no.~HB2016004.
D.Y. is supported by Postgraduate Research $\&$ Practice Innovation Program
of Jiangsu Normal University under Grant no.~2020XKT775.
\end{acknowledgments}

%
%

\begin{appendix}
\section{Distribution amplitudes}
\label{sec:app0}

In the conjugate ${\bf b}$ space of transverse momentum ${\bf k}_T$, the related $B$ meson distribution amplitude $\phi_B(x,{\bf b})$ could be given as
\beq
\phi_{B}(x,{\bf b})&=& N_{B}x^2(1-x)^2
\exp\biggl[-\frac{1}{2}\left(\frac{x m_{B}}{\omega_{B}}\right)^2
-\frac{\omega_{B}^2 {\bf b}^2}{2}\biggr] \;,
\eeq
where $\omega_B$ is the shape parameter of $\phi_B(x,{\bf b})$. $N_B$ is the normalization factor, which is correlated with the decay constant $f_{B}$ satisfying the following relation,
\beq
\int_0^1 dx \phi_{B}(x, {\bf b}=0) &=& \frac{f_{B}}{2 \sqrt{2N_c}}\;.
\label{eq:norm}
\eeq
The shape parameter $\omega_B$ has been well constrained as $\omega_B = 0.4$~GeV~\cite{Kurimoto:2001zj} for the $B_d^0$ meson and $\omega_B =0.5$~GeV~\cite{Ali:2007ff} for the $B_s^0$ meson, respectively. With the decay constants $f_{B_d^0} = 0.21$~GeV  and $f_{B_s^0} = 0.23$~GeV, the normalization factors could be obtained as $N_{B_d^0} = 101.445$ and $N_{B_s^0} = 63.67$ correspondingly. The recent developments on the $B$-meson distribution amplitude with high twists can be found in~\cite{Bell:2013tfa, Feldmann:2014ika, Braun:2017liq, Wang:2019msf, Galda:2020epp}. The effects induced by these newly developed distribution amplitudes will be left for future investigations together with highly precise measurements.

For $J/\psi$ meson, the explicit forms for the distribution amplitudes of twist-2 $\phi_{J/\psi}^{L,T}(x)$ and twist-3 $\phi_{J/\psi}^{t,v}(x)$ could be read as~\cite{Bondar:2004sv},
\beq
\phi_{J/\psi}^{L}(x) &=& \phi_{J/\psi}^{T}(x)=9.58\frac{f_{J/\psi}}{2\sqrt{2N_{c}}}x(1-x)
\biggl[\frac{x(1-x)}{1-2.8x(1-x)}\biggl]^{0.7}\;,
\label{eq:das-psi-LT}
\non
\phi_{J/\psi}^{t}(x) &=& 10.94\frac{f_{J/\psi}}{2\sqrt{2N_{c}}}(1-2x)^{2}\biggl[\frac{x(1-x)}{1-2.8x(1-x)}\biggl]^{0.7}\;,
\label{eq:da-psi-t}
\\
\phi_{J/\psi}^{v}(x) &=& 1.67\frac{f_{J/\psi}}{2\sqrt{2N_{c}}}[1+(2x-1)^{2}]\biggl[\frac{x(1-x)}{1-2.8x(1-x)}\biggl]^{0.7}\;,
\label{eq:da-psi-v}
\nonumber
\eeq
where $x$ describes the distribution of charm quark momentum in $J/\psi$ meson and $f_{J/\psi}$ is the decay constant.

And the twist-2 $\phi_{\eta_c}^v(x)$ and the twist-3 $\phi_{\eta_c}^s(x)$ distribution amplitudes for $\eta_c$ meson are collected as follows~\cite{Bondar:2004sv},
\beq
\phi_{\eta_{c}}^{v}(x) &=& 9.58\frac{f_{\eta_{c}}}{2\sqrt{2N_{c}}}x(1-x)\biggl[\frac{x(1-x)}{1-2.8x(1-x)}\biggl]^{0.7}\;,
\label{eq:da-etac-v}
\\
\phi_{\eta_{c}}^{s}(x) &=& 1.97\frac{f_{\eta_{c}}}{2\sqrt{2N_{c}}}\biggl[\frac{x(1-x)}{1-2.8x(1-x)}\biggl]^{0.7}\;,
\label{eq:da-etac-s}
\eeq
with  the $\eta_c$ decay constant  $f_{\eta_c}$ and $x$ being the momentum fraction
of charm quark in $\eta_c$ meson.

For the axial-vector meson $f_1$, the twist-2 light-cone distribution amplitudes,
i.e., $\phi_{f_q}(x)$ and $\phi_{f_q}^T(x)$, can be expanded as the Gegenbauer
polynomials~\cite{Li:2009tx}:
\beq
\phi_{f_q}(x)&=&\frac{f_{f_q}}{2\sqrt{2N_c}} 6 x  (1-x) \left[ 1  + a_{2}^\parallel\,
\frac{3}{2} ( 5(2x-1)^2  - 1 ) \right]\;,
\label{eq:da-fns-L}
\\
\phi_{f_q}^T(x)&=& \frac{f_{f_q}}{2\sqrt{2N_c}}6 x (1-x)
\left[  3 a_{1}^\perp\, (2x-1) \right] \;.
\label{eq:da-fns-T}
\eeq
And the twist-3 light-cone distribution amplitudes will be used in the following form~\cite{Li:2009tx},
\beq
\phi_{f_q}^s(x)&=&\frac{f_{f_q}}{4\sqrt{2N_c}} \frac{d}{dx}\Biggl[ 6x
(1-x) (  a_{1}^\perp (2x-1) )\Biggr]\;,
\label{eq:da-fns-s}
 \non
\phi_{f_q}^t(x)&=&\frac{f_{f_q}}{2\sqrt{2N_c}}\Biggl[
\frac{3}{2}\,a_{1}^\perp\,(2x-1) (3 (2x-1)^2-1)\Biggr]\;,
\label{eq:da-fns-t}
\\
\phi_{f_q}^v(x)&=&\frac{f_{f_q}}{2\sqrt{2N_c}}
\Biggl[\frac{3}{4} (1+(2x-1)^2) \Biggr]\;,
\label{eq:da-fns-v}
\qquad
\phi_{f_q}^a(x) = \frac{f_{f_q}}{8\sqrt{2N_c}}
\frac{d}{dx}\Biggl[6 x (1-x) \Biggr]\;,
\label{eq:da-fns-a}
\nonumber
\eeq
where $f_{f_q}$ is the decay constant of quark-flavor state $f_{q}$~\cite{Li:2009tx} and $a_2^\parallel$ and $a_1^\perp$ are the Gegenbauer moments in the $f_q$ distribution amplitudes~\cite{Yang:2007zt}.

\section{Related Functions in the PQCD approach}
\label{sec:app1}

The threshold resummation function is universal and is usually parameterized in a simplified form as~\cite{Li:2001ay}
\beq
S_{t}(x)=\frac{2^{1+2c}\Gamma(3/2+c)}{\sqrt{\pi}\Gamma(1+c)}[x(1-x)]^{c},
\eeq
with $c = 0.4\pm 0.1$ and this factor is normalized to unity.

In the calculation, the Sudakov factor for (a) and (b) in Figure.\ref{fig:fig2} is symbolled as $S_{ab}(t)$,  and one for  (c) and (d) is $S_{cd}(t)$. The expression are given as the following,
\beq
S_{ab}(t)&=&s(x_{1}P_{1}^{+},b_{1})+ s(x_{3}P_{3}^{-},b_{3})\nonumber
\\
&&+s((1-x_{3})P_{3}^{-},b_{3})
-\frac{1}{\beta_{1}}\biggl[\frac{5}{6}\ln \frac{\ln(t/\Lambda)}{-\ln(b_{1}\Lambda)}+\ln \frac{\ln(t/\Lambda)}{-\ln(b_{3}\Lambda)}\biggl],
\\
S_{cd}(t)&=&s(x_{1}P_{1}^{+},b_{1})+ s_c(x_{2}P_{2}^{+},b_{2})+s_c((1-x_{2})P_{2}^{+},b_{2})
+ s(x_{3}P_{3}^{-},b_{1})\nonumber
\\
 && +s((1-x_{3})P_{3}^{-},b_{1}) -\frac{1}{\beta_{1}}\biggl[\frac{11}{6}\ln \frac{\ln(t/\Lambda)}{-\ln(b_{1}\Lambda)}+\ln \frac{\ln(t/\Lambda)}{-\ln(m_c\Lambda)}\biggl]\;,
\eeq
where the functions $s(q,b)$ and $s_c(q,b)$ could be found in Ref.~\cite{Lu:2000em} and \cite{Liu:2018kuo}, respectively.

\end{appendix}

%
%


\begin{thebibliography}{99}

\bibitem{Cheng:2011pb}
H.~Y.~Cheng,
Phys. Lett. B \textbf{707}, 116-120 (2012).

\bibitem{Burakovsky:1997dd}
L.~Burakovsky and J.~T.~Goldman,
Phys. Rev. D \textbf{56}, R1368-R1372 (1997).

\bibitem{Chen:2015iqa}
K.~Chen, C.Q.~Pang, X.~Liu and T.~Matsuki,
Phys.\ Rev.\ D {\bf 91}, 074025 (2015).

\bibitem{ParticleDataGroup:2022pth}
R.~L.~Workman \textit{et al.} [Particle Data Group],
PTEP \textbf{2022}, 083C01 (2022).

\bibitem{HFLAV:2022pwe}
Y.~Amhis \textit{et al.} [HFLAV],
[arXiv:2206.07501 [hep-ex]].

\bibitem{Du:2022nno}
M.~C.~Du, Y.~Cheng and Q.~Zhao,
Phys. Rev. D \textbf{106}, 054019 (2022).

\bibitem{Gidal:1987bn}
G.~Gidal, J.~Boyer, F.~Butler, D.~Cords, G.S.~Abrams, D.~Amidei, A.R.~Baden and T.~Barklow {\it et al.},
Phys.\ Rev.\ Lett.\  {\bf 59}, 2012 (1987).

\bibitem{Close:1997nm}
F.E.~Close and A.~Kirk,
Z.\ Phys.\ C {\bf 76}, 469 (1997).

\bibitem{Li:2000dy}
D.M.~Li, H.~Yu and Q.X.~Shen,
Chin.\ Phys.\ Lett.\  {\bf 17}, 558 (2000).

\bibitem{Carvalho:2002fh}
W.S.~Carvalho, A.S.~de Castro and A.C.B.~Antunes,
J.\ Phys.\ A {\bf 35}, 7585 (2002).

\bibitem{Li:2005eq}
D.M.~Li, B.~Ma and H.~Yu,
Eur.\ Phys.\ J.\ A {\bf 26}, 141 (2005).

\bibitem{Yang:2007zt}
K.C.~Yang,
Nucl.\ Phys.\ B {\bf 776}, 187 (2007).

\bibitem{Cheng:2007mx}
H.Y.~Cheng and K.C.~Yang,
Phys.\ Rev.\ D {\bf 76}, 114020 (2007).

\bibitem{Yang:2008xw}
K.C.~Yang,
Phys.\ Rev.\ D {\bf 78}, 034018 (2008).

\bibitem{Cheng:2008gxa}
H.Y.~Cheng and K.C.~Yang,
Phys.\ Rev.\ D {\bf 78}, 094001 (2008)
[Erratum-ibid.\ D {\bf 79}, 039903 (2009)].

\bibitem{Yang:2010ah}
K.C.~Yang,
Phys.\ Rev.\ D {\bf 84}, 034035 (2011).

\bibitem{Dudek:2011tt}
J.J.~Dudek, R.G.~Edwards, B.~Joo, M.J.~Peardon, D.G.~Richards and C.E.~Thomas,
Phys.\ Rev.\ D {\bf 83}, 111502 (2011).

\bibitem{Stone:2013eaa}
S.~Stone and L.~Zhang,
Phys.\ Rev.\ Lett.\  {\bf 111}, 062001 (2013).

\bibitem{Dudek:2013yja}
J.J.~Dudek, R.G.~Edwards, P.~Guo and C.E.~Thomas,
Phys.\  Rev.\  D  {\bf 88}, 094505 (2013).

\bibitem{Liu:2014doa}
X.~Liu and Z.~J.~Xiao,
Phys. Rev. D \textbf{89}, 097503 (2014).

\bibitem{Liu:2014jsa}
X.~Liu, Z.J.~Xiao, J.W.~Li and Z.T.~Zou,
Phys. Rev. D \textbf{91}, 014008 (2015).

\bibitem{Close:2015rza}
F.~E.~Close and A.~Kirk,
Phys. Rev. D \textbf{91}, 114015 (2015).

\bibitem{Molina:2016pbg}
R.~Molina, M.~D\"oring and E.~Oset,
Phys. Rev. D \textbf{93}, 114004 (2016).

\bibitem{Liu:2016rqu}
X.~Liu, Z.J.~Xiao and Z.T.~Zou,
Phys. Rev. D \textbf{94}, 113005 (2016).

\bibitem{Jiang:2020eml}
Z.~Jiang, D.H.~Yao, Z.T.~Zou, X.~Liu, Y.~Li and Z.J.~Xiao,
Phys. Rev. D \textbf{102}, 116015 (2020).

\bibitem{He:2021exv}
D.~He, X.~Luo, Y.~Xie and H.~Sun,
Phys. Rev. D \textbf{103}, 094007 (2021).

\bibitem{Aaij:2013rja}
R.~Aaij {\it et al.}  [LHCb Collaboration],
Phys.\ Rev.\ Lett.\  {\bf 112}, 091802 (2014).

\bibitem{Cheng:2000kt}
H.Y.~Cheng and K.C.~Yang,
Phys. Rev. D \textbf{63}, 074011 (2001).

\bibitem{Cheng:2001ez}
H.Y.~Cheng, Y.Y.~Keum and K.C.~Yang,
Phys. Rev. D \textbf{65}, 094023 (2002).

\bibitem{Song:2002gw}
Z.z.~Song, C.~Meng and K.T.~Chao,
Eur. Phys. J. C \textbf{36}, 365 (2004).

\bibitem{Meng:2005er}
C.~Meng, Y.J.~Gao and K.T.~Chao,
Phys. Rev. D \textbf{87}, 074035 (2013).

\bibitem{Chen:2005ht}
C.H.~Chen and H.-n.~Li,
Phys.\ Rev.\ D {\bf 71}, 114008 (2005).

\bibitem{Li:2006vq}
H.-n.~Li and S.~Mishima,
\jhep \textbf{03}, 009 (2007).

\bibitem{Li:2007xf}
J.W.~Li and D.S.~Du,
Phys. Rev. D \textbf{78}, 074030 (2008).

\bibitem{Beneke:2008pi}
M.~Beneke and L.~Vernazza,
Nucl. Phys. B \textbf{811}, 155-181 (2009).

\bibitem{Liu:2009yno}
X.~Liu, Z.Q.~Zhang and Z.J.~Xiao,
Chin. Phys. C \textbf{34}, 937 (2010).

\bibitem{Colangelo:2010wg}
P.~Colangelo, F.~De Fazio and W.~Wang,
Phys. Rev. D \textbf{83}, 094027 (2011).

\bibitem{Liu:2012ib}
X.~Liu, H.-n.~Li and Z.J.~Xiao,
Phys. Rev. D \textbf{86}, 011501 (2012).

\bibitem{Li:2012sw}
J~W.~Li, D.S.~Du and C.D.~L\"u,
Eur. Phys. J. C \textbf{72}, 2229 (2012).

\bibitem{Wang:2015uea}
W.F.~Wang, H.-n.~Li, W.~Wang and C.D.~L\"u,
Phys. Rev. D \textbf{91}, 094024 (2015).

\bibitem{Liu:2019ymi}
X.~Liu, Z.T.~Zou, Y.~Li and Z.J.~Xiao,
Phys. Rev. D \textbf{100}, 013006 (2019).

\bibitem{Liu:2013nea}
X.~Liu, W.~Wang and Y.~Xie,
Phys.\ Rev.\ D {\bf 89}, 094010 (2014).

\bibitem{Xiao:2019mpm}
Z.J.~Xiao, D.C.~Yan and X.~Liu,
Nucl. Phys. B \textbf{953}, 114954 (2020).

\bibitem{Keum:2000ph}
Y.Y.~Keum, H.-n.~Li and A.I.~Sanda,
Phys.\ Lett.\ B {\bf 504}, 6 (2001).

\bibitem{Lu:2000em}
C.D.~L\"u, K.~Ukai and M.Z.~Yang,
Phys.\ Rev.\ D {\bf 63}, 074009 (2001).

\bibitem{Ali:2007ff}
A.~Ali, G.~Kramer, Y.~Li, C.D.~L\"u, Y.L.~Shen, W.~Wang and Y.M.~Wang,
Phys.\ Rev.\ D {\bf 76}, 074018 (2007).

\bibitem{Liu:2018kuo}
X.~Liu, H.-n.~Li and Z.J.~Xiao,
Phys. Rev. D \textbf{97}, 113001 (2018).

\bibitem{Liu:2020upy}
X.~Liu, H.~n.~Li and Z.~J.~Xiao,
Phys. Lett. B \textbf{811}, 135892 (2020).

\bibitem{Abe:2001oa}
K.~Abe {\it et al.}  [Belle Collaboration],
Phys.\ Rev.\ Lett.\  {\bf 87}, 161601 (2001).

\bibitem{Buchalla:1995vs}
G.~Buchalla, A.~J.~Buras and M.~E.~Lautenbacher,
Rev. Mod. Phys. \textbf{68}, 1125-1144 (1996).

\bibitem{Zou:2015iwa}
Z.T.~Zou, A.~Ali, C.D.~L\"u, X.~Liu and Y.~Li,
Phys. Rev. D \textbf{91}, 054033 (2015).

\bibitem{Keum:2000wi}
Y.Y.~Keum, H.-n.~Li and A.I.~Sanda,
Phys.\ Rev.\ D {\bf 63}, 054008 (2001).

\bibitem{Lu:2002ny}
C.D.~L\"u and M.Z.~Yang,
Eur. Phys. J. C \textbf{28}, 515 (2003).

\bibitem{Botts:1989nd}
J.~Botts and G.~F.~Sterman,
Phys. Lett. B \textbf{224}, 201 (1989).
[erratum: Phys. Lett. B \textbf{227}, 501 (1989)]

\bibitem{Li:1992nu}
H.-n.~Li and G.F.~Sterman,
Nucl. Phys. B \textbf{381}, 129-140 (1992).

\bibitem{Li:2001ay}
H.-n.~Li,
Phys. Rev. D \textbf{66}, 094010 (2002).

\bibitem{Li:2002mi}
H.-n.~Li and K.~Ukai,
Phys. Lett. B \textbf{555}, 197 (2003).

\bibitem{Li:2003yj}
H.-n.~Li,
Prog. Part. Nucl. Phys. \textbf{51}, 85 (2003).

\bibitem{Cheng:2020fcx}
S.~Cheng and Z.J.~Xiao,
Front. Phys. (Beijing) \textbf{16}, 24201 (2021).

\bibitem{Hua:2020usv}
J.~Hua, H.~n.~Li, C.~D.~Lu, W.~Wang and Z.~P.~Xing,
Phys. Rev. D \textbf{104}, 016025 (2021).

\bibitem{Bondar:2004sv}
A.~E.~Bondar and V.~L.~Chernyak,
Phys.\ Lett.\ B {\bf 612}, 215 (2005).

\bibitem{Li:2009tx}
R.H.~Li, C.D.~L\"u and W.Wang,
Phys.\ Rev.\ D {\bf 79}, 034014 (2009).

\bibitem{Verma:2011yw}
R.C.~Verma,
J. Phys. G \textbf{39}, 025005 (2012).

\bibitem{Xiao:2008sw}
Z.J.~Xiao, Z.Q.~Zhang, X.~Liu and L.B.~Guo,
Phys. Rev. D \textbf{78}, 114001 (2008).

\bibitem{Wolfenstein:1983yz}
L.~Wolfenstein,
Phys. Rev. Lett. \textbf{51}, 1945 (1983).

\bibitem{Belle-II:2018jsg}
E.~Kou \textit{et al.} [Belle-II],
PTEP \textbf{2019}, 123C01 (2019).
[erratum: PTEP \textbf{2020}, 029201 (2020)]

\bibitem{Li:2012nk}
H.-n.~Li, Y.L.~Shen and Y.M.~Wang,
Phys.\ Rev.\ D {\bf 85}, 074004 (2012).

\bibitem{Cheng:2014fwa}
S.~Cheng, Y.Y.~Fan, X.~Yu, C.D.~L\"u, and Z.J.~Xiao,
Phys.\ Rev.\ D\ {\bf 89}, 094004 (2014).

\bibitem{BaBar:2007bpi}
B.~Aubert \textit{et al.} [BaBar],
Phys. Rev. Lett. \textbf{99}, 201802 (2007).

\bibitem{LHCb:2013vga}
R.~Aaij \textit{et al.} [LHCb],
Phys. Rev. D \textbf{88}, 052002 (2013).

\bibitem{LHCb:2011aa}
R.~Aaij \textit{et al.} [LHCb],
Phys. Rev. Lett. \textbf{108}, 101803 (2012).

\bibitem{Beneke:2006hg}
M.~Beneke, J.~Rohrer and D.~Yang,
Nucl.\ Phys. B {\bf 774}, 64 (2007).

\bibitem{Kurimoto:2001zj}
T.~Kurimoto, H.-n.Li and A.I.~Sanda,
Phys.\ Rev.\ D {\bf 65}, 014007 (2002).

\bibitem{Bell:2013tfa}
G.~Bell, T.~Feldmann, Y.~M.~Wang and M.~W.~Y.~Yip,
\jhep \textbf{11}, 191 (2013).

\bibitem{Feldmann:2014ika}
T.~Feldmann, B.O.~Lange and Y.M.~Wang,
Phys.\ Rev.\ D {\bf 89}, 114001 (2014).

\bibitem{Braun:2017liq}
V.M.~Braun, Y.~Ji and A.N.~Manashov,
\jhep {\bf 05}, 022 (2017).

\bibitem{Wang:2019msf}
W.~Wang, Y.M.~Wang, J.~Xu and S.~Zhao,
Phys.\ Rev.\ D {\bf 102}, 011502 (2020).

\bibitem{Galda:2020epp}
A.M.~Galda and M.~Neubert,
Phys.\ Rev.\ D {\bf 102}, 071501 (2020).




\end{thebibliography}
\end{document}